\documentclass{aa}
\usepackage{times}
\usepackage{graphics}
\usepackage{xspace}
\usepackage{epsfig}
\usepackage{jwaabib}
\usepackage{rotating}
\usepackage{dcolumn}

\newcommand{\myrule}{\rule[-0.2cm]{0.cm}{0.4cm}}

\newcommand{\yygem}{YY\,Gem\xspace}

\begin{document}


\title{Simultaneous X-ray spectroscopy of \yygem \\ with {\em Chandra} and {\em XMM-Newton}}

\author{B. Stelzer\inst {1} 
\and V. Burwitz\inst {1} 
\and M. Audard\inst{2}
\and M. G\"udel\inst{2}
\and J.-U. Ness\inst{3} 
\and N. Grosso\inst{1}
\and R. Neuh\"auser\inst {1} 
\and J. H. M. M. Schmitt\inst {3}
\and P. Predehl\inst {1}
\and B. Aschenbach\inst{1}
}

\institute{Max-Planck-Institut f\"ur extraterrestrische Physik,
  Postfach 1312, 
  D-85741 Garching,
  Germany \and
Paul Scherrer Institut, W\"urenlingen \& Villigen, 5232 Villigen PSI,
Switzerland \and
Hamburger Sternwarte, Gojenbergsweg 112, D-21029 Hamburg, Germany} 

\offprints{B. Stelzer}
\mail{B. Stelzer, stelzer@xray.mpe.mpg.de}
\titlerunning{X-ray spectroscopy of \yygem}

\date{Received $<$25-03-02$>$ / Accepted $<$20-06-02$>$}

\abstract{
We report on a detailed study of the X-ray spectrum of the nearby eclipsing
spectroscopic binary \yygem. Observations were obtained simultaneously
with both large X-ray observatories, {\em XMM-Newton} and {\em Chandra}.
We compare the high-resolution spectra acquired with the Reflection Grating
Spectrometer onboard {\em XMM-Newton} and with the Low Energy Transmission
Grating Spectrometer onboard {\em Chandra}, 
and evidence in direct comparison the good 
performance of both instruments in terms of wavelength and flux calibration.
The strongest lines
in the X-ray spectrum of \yygem are from oxygen. Oxygen line ratios 
indicate the presence of a low-temperature component ($1-4$\,MK) 
with density $n_{\rm e} \leq 2\,10^{10}\,{\rm cm^{-3}}$.
The X-ray lightcurve reveals two flares and a dip corresponding to the 
secondary eclipse.
An increase of the density during phases of high activity is suggested
from time-resolved spectroscopy.
Time-resolved global fitting of the European Photon Imaging Camera 
CCD spectrum traces the evolution of temperature and emission measure during 
the flares. These medium-resolution spectra show that temperatures 
$> 10^7$\,K are relevant in the corona of YY\,Gem although not as dominant  
as the lower temperatures represented by the strongest lines in the 
high-resolution spectrum. 
Magnetic loops with length on the order of $10^9$\,cm,  
i.e., about 5\,\% of the radius of each star,
are inferred from a comparison with a 
one-dimensional hydrodynamic model. This suggests that the flares did not
erupt in the (presumably more extended) inter-binary magnetosphere 
but are related to one of the components of the binary.
}

\maketitle

\section{Introduction}\label{sect:intro}

The Castor system comprises three visual stars, all of which
are spectroscopic binaries. The optically faintest of the three
components, \yygem (= Castor C), is of major importance to stellar
evolution studies: 
being an eclipsing spectroscopic binary ($i \sim 86^\circ$;
\cite{Pettersen76.1}) very near or on the main sequence it
allows for the determination of masses and radii of both components, and
for a test of evolutionary tracks. Both stars in the \yygem system are
of nearly equal spectral type, dM1e, in a synchronous orbit with a 
period of 0.81\,d. \yygem was the first late-type star on which
periodic photometric variability was detected (\cite{Kron52.1}). 
\citey{Bopp73.1} have explained similar brightness variations on the
flare star BY\,Dra by rotation of star spots, and stars displaying this
phenomenon are since then termed BY\,Dra variables. Doppler images of \yygem
have revealed spots at mid-latitudes on both stars (\cite{Hatzes95.1}).   

X-rays from the Castor sextuplet were first recorded by the
{\em Einstein} satellite (\cite{Vaiana81.1}, \cite{Caillault82.1}, 
\cite{Golub83.1}). 
Since then the system was observed by virtually all X-ray observatories;
see e.g., \citey{Pallavicini90.1}, \citey{Gotthelf94.1},
\citey{Schmitt94.1}, \citey{Guedel01.1}. 
As the spatial resolution of the instruments improved, X-ray emission 
from individual members of the Castor system could be identified. 
At present, all three visual binaries in the sextuplet are known to be 
X-ray sources.
The X-ray emission from Castor\,A and~B
is commonly attributed to their late-type companions 
because the primaries, being early A-type stars without a 
convective envelope 
and consequently with no ability to support the dynamo mechanism,
are not expected to be X-ray sources.

In nearly all of the previous X-ray studies of \yygem 
flare activity was reported. The luminosity of \yygem in the soft 
X-ray band ranges 
between $\sim (2 - 8)\,10^{29}\,{\rm erg/s}$.
Flares have been observed also in other parts of its 
electromagnetic spectrum. \citey{Doyle90.1} found that the flare activity 
on \yygem is more than an order
of magnitude larger during out-of-eclipse times as compared to times
when one of the stars is eclipsed. On the basis of these U band
observations it was suggested that the amount of
magnetic energy is largest in the inter-binary space, leading to frequent
energy release in this region. A periodicity of the optical flaring rate
was reported by \citey{Doyle90.2}, but has not been confirmed so far.  
The chromosphere and transition region was examined by 
\citey{Haisch90.1}, and moderate variability throughout the orbital cycle
was detected in the emission lines. 

The extraordinary variability of \yygem has instigated us to 
carry out a coordinated multi-wavelength campaign comprising radio, optical
and X-ray observations. 
As a unique example of an eclipsing spectroscopic binary with nearly
identical components \yygem is ideally suited for a study of stellar 
coronal structure. In particular we were aiming at looking into the
possibility of enhanced flare activity in the inter-binary space as
suggested by \citey{Doyle90.1}. Magnetic structures in between
the components of binaries have been discussed also for RS\,CVn systems 
(e.g., \cite{Uchida83.1}).
Additional information about the coronal geometry 
can be obtained from observations of X-ray eclipses. 
The depth and duration of the eclipses provide a measure for the spatial 
extent of the corona and contribution of the two components in the 
binary. From a 3D-deconvolution of the X-ray lightcurve of \yygem
(using an earlier observation by {\em XMM-Newton}) \citey{Guedel01.1}
found that the coronae of both stars in the binary are inhomogeneously
structured with brighter areas at mid-latitudes.
Obscuration of coronal features can also be used to localize and
constrain the emitting region in active stars
(see e.g., Gunn et~al. 1997, 1999, \cite{Schmitt99.1}). 

As part of our multi-wavelength project the Castor system 
was observed simultaneously with both {\em XMM-Newton} and {\em Chandra}. 
While in the EPIC MOS image presented by \citey{Guedel01.1} Castor~A and~B
were spatially resolved, 
our {\em Chandra} observation allows for the
first time to separate the X-ray spectra of Castor~A and~B 
from each other. We will examine the X-ray characteristics of Castor~A and~B 
in a related publication (Stelzer et al., in prep.). 
Here we concentrate on the intermediate and
high-resolution X-ray spectrum of YY\,Gem. 
We emphasize that this observation presents an excellent opportunity
to cross-check the calibration of the dispersive instruments onboard both
satellites. We analyse the emission line spectrum observed with the 
{\em Chandra} Low-Energy Transmission Grating Spectrometer (LETGS) 
and the Reflection Grating Spectrometer (RGS) onboard {\em XMM-Newton} 
independently of each other, and derive observed line fluxes for both 
instruments.
Emission lines in the spectral region of the LETGS and RGS include the
helium-like triplets from C\,V to Si\,XIII, the Lyman series of 
hydrogen-like ions, and numerous iron L-shell transitions. 
By means of time-resolved analysis of spectral parameters
we check for spectral variability related to 
changes in the activity level of YY\,Gem.
We use both `global' fits to the EPIC X-ray spectrum from $0.3$ to $10$\,keV
and modeling of individual emission lines in the range 2-175\,\AA~~
(resolved by the gratings onboard both {\em Chandra} and {\em XMM-Newton}) to
infer the time evolution of temperature and density of the emitting plasma.

The {\em XMM-Newton} and {\em Chandra} observations are introduced in 
Sect.~\ref{sect:obs}. In Sect.~\ref{sect:highres-spectra} we present
results from high-resolution spectroscopy, involving a detailed comparison
of the contemporaneous LETGS and RGS spectrum. 
Time-resolved spectroscopy is discussed in Sect.~\ref{sect:timeres}.
We draw conclusions on the coronal structure by means of loop modeling 
(Sect.~\ref{sect:loops}), and summarize the results in 
Sect.~\ref{sect:conclusions}.

\section{Observations}\label{sect:obs}

\begin{figure*}
\begin{center}
\resizebox{19cm}{!}{\includegraphics{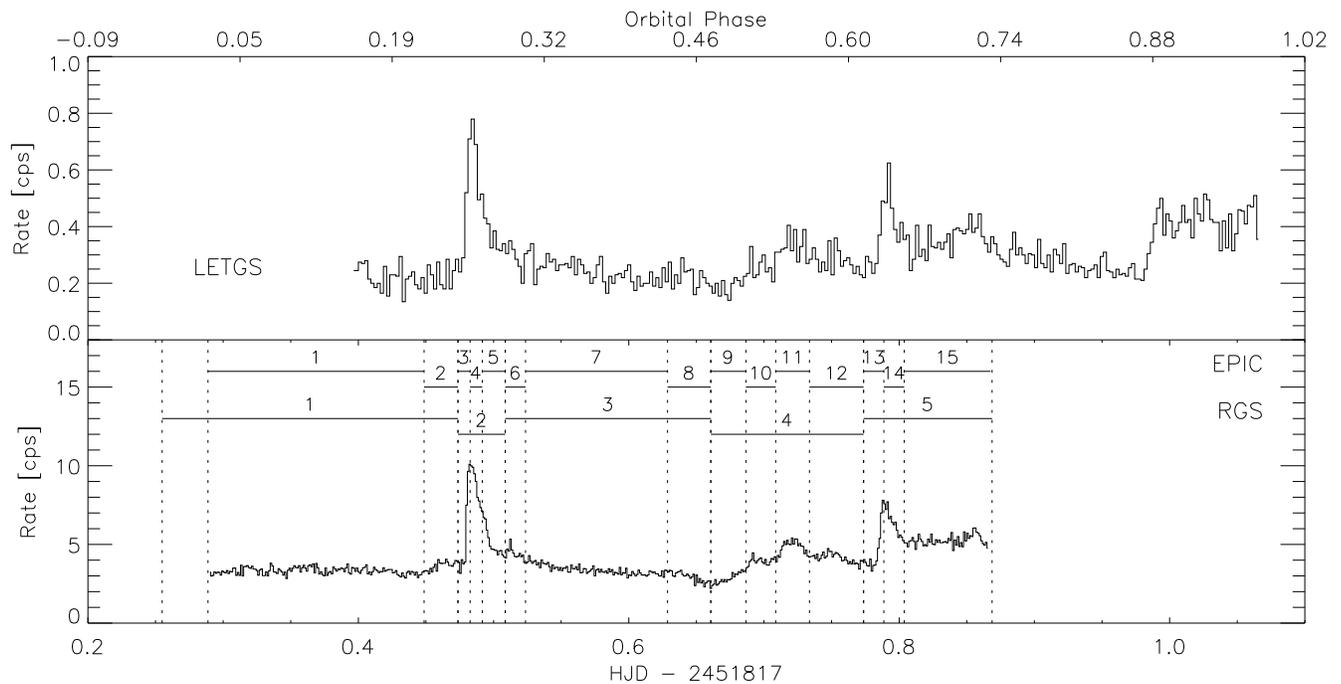}}
\caption{Background subtracted X-ray lightcurve of \yygem on 2000 Sep 29/30. 
{\em top} - {\em Chandra} LETG zeroth order with 200\,s binsize, 
{\em bottom} - {\em XMM-Newton} EPIC pn with 100\,s binsize. 
The orbital phase was computed with 
the ephemeris by \protect\citey{Torres02.1}. 
The time intervals selected for time-resolved spectroscopy are indicated 
by horizontal lines above the lightcurves.} 
\label{fig:timeintervals}
\end{center}
\end{figure*}

\yygem was observed by both {\em Chandra} (Obs-ID 28) and 
{\em XMM-Newton} (Obs-ID 0112880801) on
Sep 29/30 2000 for a total observing time of 59\,ksec and 55\,ksec,
respectively. Both observations were part of the Guaranteed Time program. 
X-ray lightcurves obtained with the individual instruments demonstrating
the times of observation are shown in Fig.~\ref{fig:timeintervals}.
%
%

We have computed the orbital phases for the time of observation using
the ephemeris of \citey{Torres02.1}. 
Two large flare events, the secondary eclipse, a flare-like
feature after secondary eclipse, and an extended ``high state'' at the
end of the {\em Chandra} observation are seen. The structure of 
the X-ray lightcurve resembles that of the earlier {\em XMM-Newton} 
observation of \yygem made in April 2000 (\cite{Guedel01.1}). 
The larger of the two
flares and the feature after secondary eclipse are seen at similar 
orbital phases in both observations.

\subsection{{\em XMM-Newton}}\label{subsect:xmm}

The {\em XMM-Newton} observation on Sep 29/30, 2000 
was carried out in full-frame mode of both EPIC instruments and 
with the thick filter inserted for both pn and MOS to avoid
optical loading from Castor. 
The Optical Monitor was in closed position due to the optical
brightness of the Castor system. 
For a description of the X-ray instruments onboard {\em XMM-Newton}
see \citey{Jansen01.1}, \citey{denHerder01.1}, \citey{Strueder01.1}, 
\citey{Turner01.1}, and \citey{Aschenbach02.1}.

For the analysis of the EPIC spectrum source photons were 
selected from a circular region centered on the position of \yygem.
An appropriate outer radius for the extraction of source photons is 
the value where the background subtracted 
radial distribution of integrated counts flattens out, in this case
$\sim 26^{\prime\prime}$.
With this choice we make sure that 
the data are not contaminated by contributions from Castor\,AB (at a 
separation of $74^{\prime\prime}$ from \yygem). 
A background spectrum was extracted from a nearby position with the same
area as the source extraction region.
The EPIC data were analysed with the {\em XMM-Newton} Science Analysis
Software (SAS) version from May 2000. 
We used the pn detector response made available by the hardware team
in April 2001 and the MOS responses released in February 2001
(see Sect.~\ref{subsubsect:pile-up} for more details).
The RGS data were analysed with the SAS version 5.2. CCD PI 
filtering allows to separate the orders. We concentrate on the first 
order spectrum. The background spectrum is obtained from a region offset
from the spectrum of YY\,Gem and Castor\,AB 
in the cross-dispersion direction.

\subsection{{\em Chandra}}\label{subsect:chandra}

With {\em Chandra} we used the LETGS combining the Low-Energy Transmission
Grating (LETG) with the High Resolution Camera (HRC) (see
\cite{Brinkman00.1}).
The {\em Chandra} observation began  $\sim 2.5$\,h after
the start of the pointing with {\em XMM-Newton}. After the
{\em XMM-Newton} observation ended {\em Chandra} continued
to observe for $\sim 5$\,h (see Fig.~\ref{fig:timeintervals}).

The data obtained from the {\em Chandra} science center
was reprocessed following the CIAO 2.1 science 
threads\footnote{http://asc.harvard.edu/ciao/documents-threads.html} for
applying the newest degap-corrections to the HRC-S data as well as
generating
new event level 2 files, in which background is reduced significantly by a
spatially dependent ``light pha filtering''  described in the thread.

The spectra were then extracted from these event files
using IDL routines. For the source we used the `bow tie' extraction
region as described in the
{\em Chandra} Proposers' 
Observatory Guide\footnote{http://asc.harvard.edu/udocs/docs/docs.html} (POG).
Large areas above and below the source extraction region were selected
for the background, such that the ratio between source and background 
area is 7.5 (see Fig.~1 in \cite{Burwitz01.1} for an illustration).
Towards longer wavelengths $\lambda $\,$>$\,48.4\,\AA\ the
background regions become wider in order to maintain the
constant ratio of areas between source and background spectral
bins.

\section{High-resolution Spectroscopy}\label{sect:highres-spectra}

In this section we present and compare results from the
high-resolution dispersive instruments on {\em Chandra} and {\em XMM-Newton}.
We took care that the Castor~AB components
are excluded from the background extraction regions.
Lists of all identified lines are given 
in Tables~\ref{tab:rgs_letgs} and~\ref{tab:letgs}. 
We derive photon fluxes and
use line ratios for coronal plasma diagnostics. In this context the 
triplets of He-like ions comprising the resonance line $r$, the
intercombination line $i$, and the forbidden line $f$, 
are of particular interest as their intensity ratios 
$R = f/i$ and $G=(f+i)/r$ provide an estimate for the electron 
density $n_{\rm e}$ and electron temperature $T_{\rm e}$ 
of the respective lines.
The temperature sensitivity of $G$ is due to the collisional excitation
rates which have different temperature dependence for the resonance line
compared to the forbidden and intercombination line. The $G$ ratio also
indicates whether the plasma is dominated by collisions or
photo-ionization: a strong $r$ line corresponds to 
collisionally dominated plasmas 
($G \sim 1$), while $r$ is weak in photo-ionized environments with $G > 4$  
(see \cite{Porquet00.1}). 
We caution that a given line does actually form over an extended range 
of temperatures in a non-isothermal plasma, and thus the temperature derived
from $G$ may be misleading.
The number of collisions increases with density.
As a consequence the upper level of the $f$ transition is depopulated 
and fills the upper level of the $i$ transition. This results in a decrease
of $R$ with increasing $n_{\rm e}$. Only in the low- and high-density
limits is $R$ independent of $n_{\rm e}$.

\subsection{Line identifications}\label{subsect:lineiden}

\begin{figure*}
\begin{center}
\hspace*{-1.7cm}\resizebox{20cm}{!}{\includegraphics{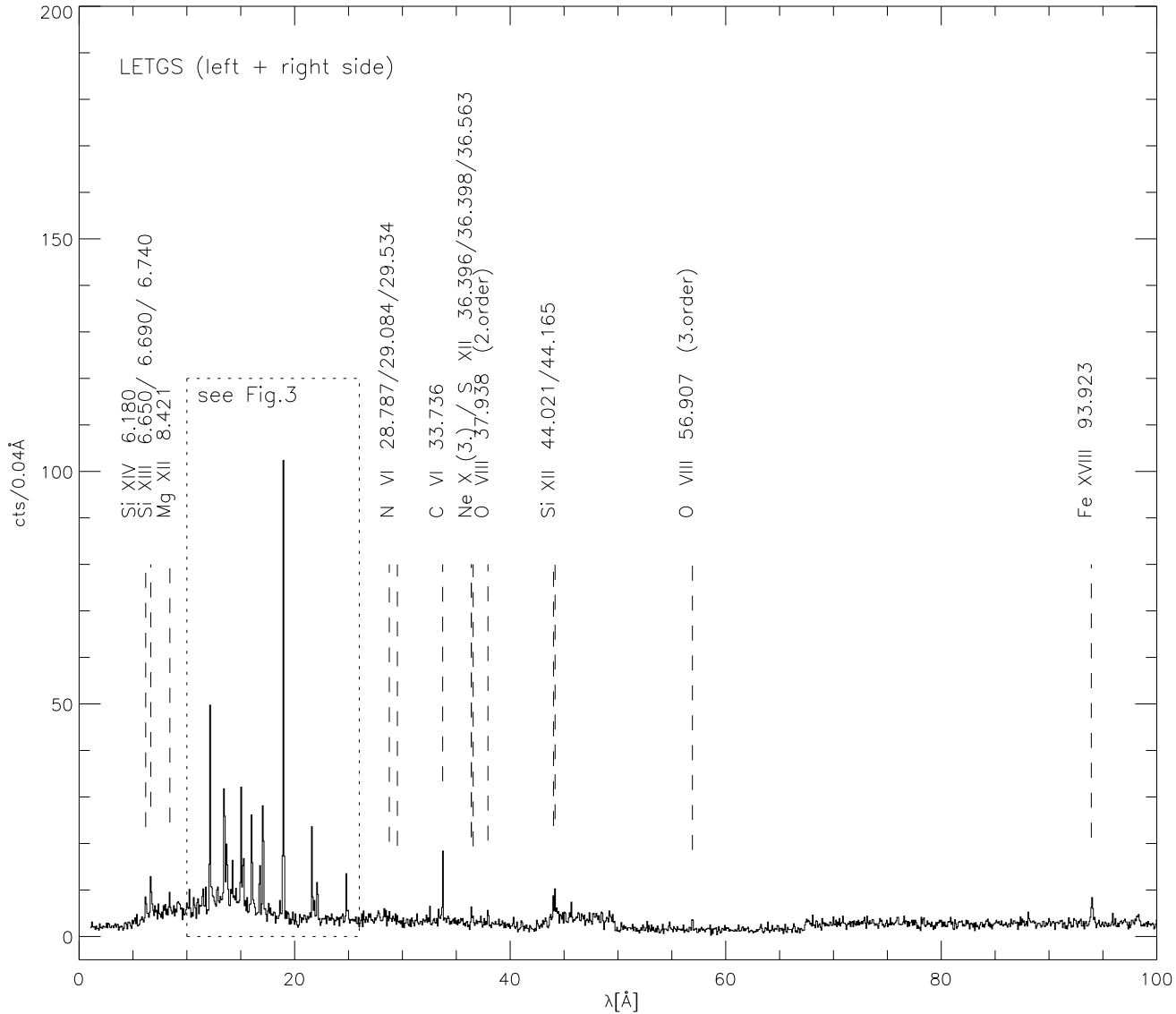}}
\caption{Time-averaged {\em Chandra} LETGS spectrum 
(left and right side added); total exposure time is 58\,ksec, 
and binsize is 0.04\,\AA. 
Only few emission lines are identified above 100\,\AA~ 
(108.4\,\AA, 117.2\,\AA, 128.73\,\AA, and 132.9\,\AA; not displayed here).
The region within the dotted rectangle ($10-26$\,\AA~) 
is shown in more detail in Fig.~\ref{fig:letgs_rgs_spec}.}
\label{fig:letgs_spec}
\end{center}
\end{figure*}

\begin{sidewaysfigure*}
\begin{center}
\vspace*{-2cm}\hspace*{-1cm}\resizebox{27cm}{!}{\includegraphics{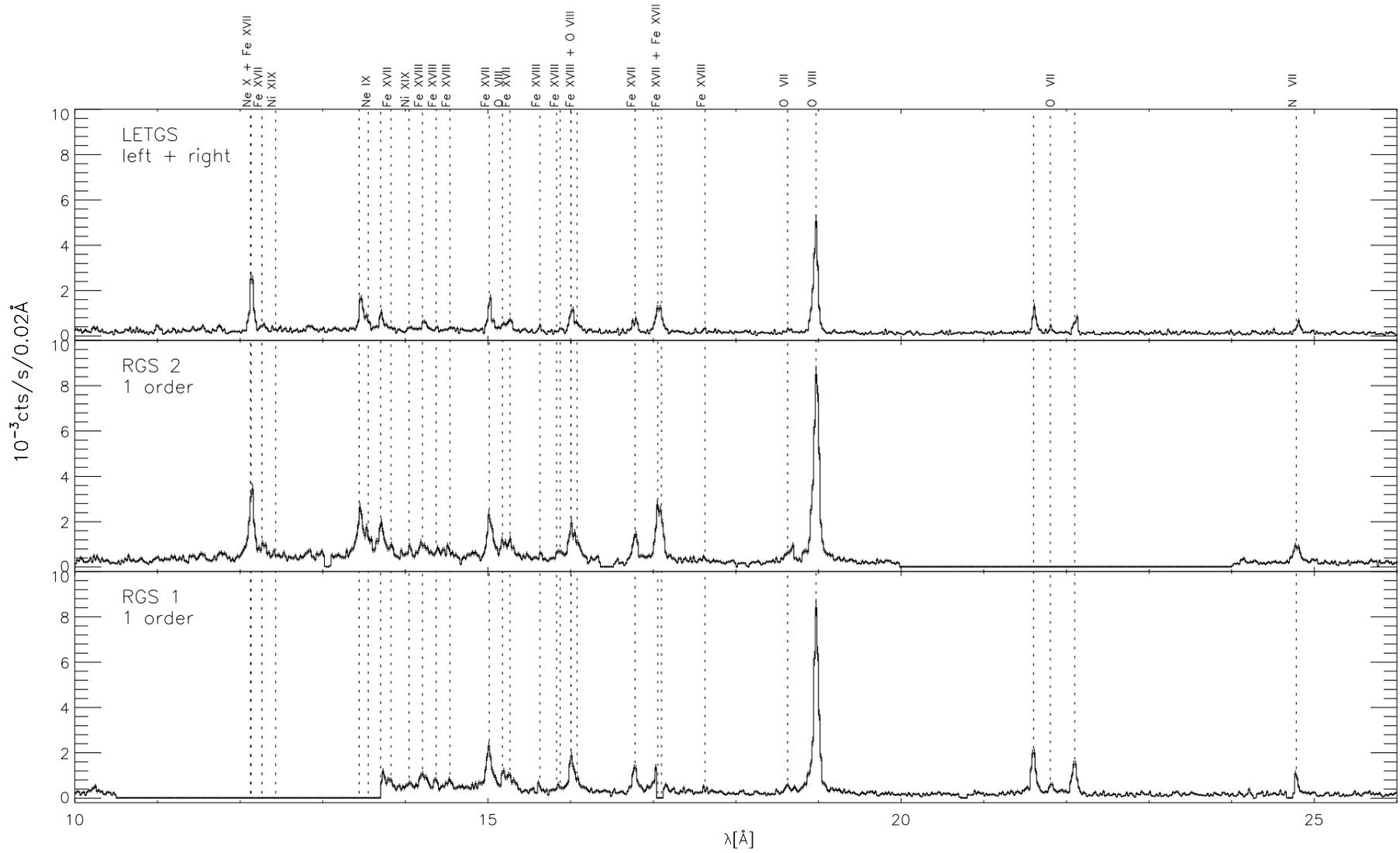}}
\end{center}
\caption{Comparison of the contemporaneous {\em XMM-Newton} RGS\,1 and
RGS\,2, and {\em Chandra} LETGS spectra 
of \yygem in the range $\lambda = 10-26$\,\AA. 
Straight horizontal lines in the RGS spectra represent gaps
due to CCD chain failure or individual chip separation.}
\label{fig:letgs_rgs_spec}
\end{sidewaysfigure*}

We extracted and analysed the RGS and the LETGS 
spectra of \yygem as described in Sect.~\ref{subsect:xmm}
and~\ref{subsect:chandra}. The time-averaged LETGS spectrum in the wavelength 
range from $0-100$\,\AA~ is displayed in Fig.~\ref{fig:letgs_spec}. 
We have added left (negative) and right (positive) side of the spectrum.
%
Above 100\,\AA~ only few lines are seen, all from highly ionized iron.
A close-up of the region with the strongest emission 
lines is shown in Fig.~\ref{fig:letgs_rgs_spec} together with the 
time-averaged first order RGS spectrum in the same wavelength region. 
We show the spectrum in units of cts/s/bin. Since the RGS and the LETGS 
observations overlap for about $75$\,\% in time, the measured line
fluxes should be similar. 
The count rates are higher for RGS than for LETGS 
throughout most of the displayed spectral region, and demonstrate 
directly the higher sensitivity of the RGS, 
except at wavelengths $\leq 10\,$\AA.
The full wavelength range of the RGS extends from $6-35$\,\AA. However,
the only lines found in the RGS spectrum of \yygem outside the range displayed
in Fig.~\ref{fig:letgs_rgs_spec} are 
the silicon lines between $6-7$\,\AA, 
the resonance line of the triplet of He-like N\,VI, 
and transitions from H-like carbon.
%
%

We identified the most prominent lines in both the LETGS and the
RGS spectrum. The line centers and fluxes were determined with the 
CORA\footnote{CORA can be downloaded from http://www.hs.uni-hamburg.de/DE/Ins/Per/Ness/Cora} 
line fitting application (\cite{Ness01.1}).
CORA was designed for low count rate spectra, based on Poisson statistics 
and employs a maximum likelihood method to determine the best fit parameters. 
The background is treated as a constant
in the vicinity of individual lines. 
For the LETGS spectrum we used Gaussians to approximate the line profile. 
To avoid artificial broadening of lines resulting from uncertainties in the
wavelength calibration, the positive and negative sides of the spectrum 
were analysed separately. 
Then we checked the agreement, and if adequate, we merged the 
positive and negative sides to increase the signal 
for the analysis of line ratios. 
For RGS Lorentzians represent a better approximation to the 
instrumental profile. To improve the S/N the first order spectra of RGS\,1 
and 2 were co-added. 
Both RGS\,1 and RGS\,2 have a $\sim 3-4$\,\AA~ wide 
gap in the spectrum as a consequence of CCD chain failure 
($10.4-13.8$\,\AA~ for RGS\,1 and $19.8-24.2$\,\AA~ for RGS\,2), 
and further small ($\sim 0.1$\,\AA) gaps due to the finite separation of
adjacent chips. 

The line parameters derived from LETGS and RGS are summarized in 
Tables~\ref{tab:rgs_letgs} and~\ref{tab:letgs}. 
The results from the positive and negative side of the LETGS spectrum
are listed separately to examine the accuracy of the wavelength calibration.
We find that the line centers are in excellent agreement.
Apart from the observed 
line center and width Tables ~\ref{tab:rgs_letgs} and~\ref{tab:letgs} present
the total number of counts $I$, 
the photon flux, and the identification and predicted line
positions adopted from \citey{Mewe85.1}, \citey{Mewe95.1}, and 
\citey{Phillips99.1}. 
We used the effective areas of the respective instrument to compute 
the photon flux. 
The photon flux does not depend on instrumental properties and can be used
for direct comparison of LETGS and RGS. 
%
%
\begin{figure}
\begin{center}
\resizebox{9cm}{!}{\includegraphics{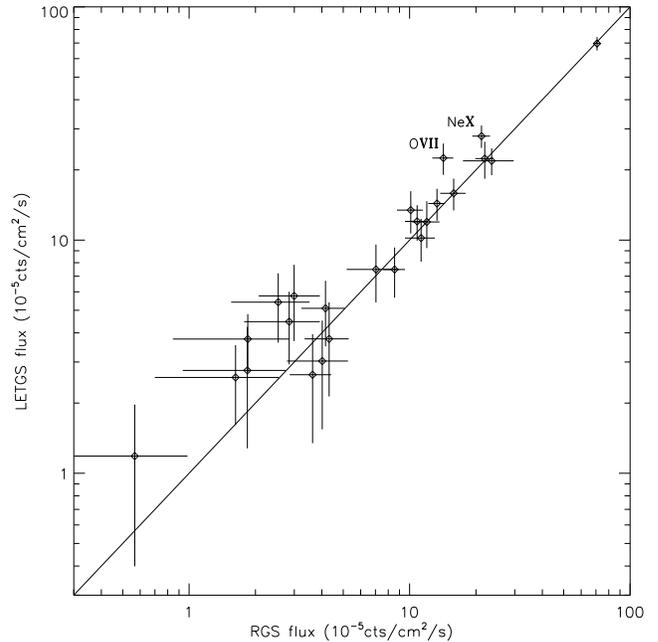}}
\caption{Comparison of LETGS and RGS fluxes for individual emission lines; 2\,$\sigma$
uncertainties.}
\label{fig:fluxcomp_rgs_letgs}
\end{center}
\end{figure}

In Fig.~\ref{fig:fluxcomp_rgs_letgs} we present a comparison of
the LETGS and RGS fluxes for individual lines. Our simultaneous observations 
allow to demonstrate that the measurements by these
two instruments are consistent.
Most photon fluxes are in agreement within one or two sigma.
The largest deviations between LETGS and RGS occur for the O\,VII triplet
and the Ne\,X Ly$\alpha$ line.
The deviations in O\,VII can most likely be attributed to an oxygen 
absorption edge in the RGS which reduces the effective area at the position 
of the O\,VII
triplet, and had not yet been corrected for in the response matrix
at the time of analysis 
(\cite{denHerder02.1}). The discrepancy in Ne\,X seems to result from a
problem in the positive side of the LETGS, as its negative side and RGS
provide comparable fluxes. Note that the photon fluxes in 
Tables~\ref{tab:rgs_letgs} and~\ref{tab:letgs}
denote the combined flux of both unresolved stars in the \yygem binary.

%
%
%
\begin{sidewaystable*}\scriptsize
\begin{center}
\caption{Results from line fitting of the time averaged first order LETGS
and RGS spectra. Positive and negative side of the LETGS are listed
separately to demonstrate the accuracy of the wavelength calibration. 
RGS\,1 and RGS\,2 spectra were combined except for lines which fall
in the wavelength ranges corresponding to failed CCDs or gaps between chips.
Line blends are indicated by quotation marks ($^{\prime\prime}$). 
Lines which are identified but not properly resolved are marked by asterisks 
($^*$).
For RGS all lines are represented by Lorentzian profiles, and for LETGS by
Gaussians. 
The typical error on the line width is $\sim 0.005$\,\AA. Line identifications
are from \protect\citey{Mewe85.1}, \protect\citey{Mewe95.1}, and 
\protect\citey{Phillips99.1}.
Lines detected only by LETGS are listed in Table~\ref{tab:letgs}.}
\label{tab:rgs_letgs}
\begin{tabular}{rrrr|rrrr|rrrr|lll} \hline
\multicolumn{1}{c}{$\lambda$} \myrule & \multicolumn{1}{c}{$\sigma$} & \multicolumn{1}{c}{I} & \multicolumn{1}{c|}{Photon Flux$^a$} & \multicolumn{1}{c}{$\lambda$} \myrule & \multicolumn{1}{c}{$\sigma$} & \multicolumn{1}{c}{I} & \multicolumn{1}{c|}{Photon Flux$^a$} & \multicolumn{1}{c}{$\lambda$} & \multicolumn{1}{c}{$\sigma$} & \multicolumn{1}{c}{I} & \multicolumn{1}{c|}{Photon Flux$^a$} & \multicolumn{3}{c}{Identification} \\
\multicolumn{1}{c}{[\AA]} \myrule & \multicolumn{1}{c}{[\AA]} & \multicolumn{1}{c}{[cts]} & \multicolumn{1}{c|}{[$10^{-5}\,\frac{\rm ph}{\rm s\,cm^2}$]} & \multicolumn{1}{c}{[\AA]} \myrule & \multicolumn{1}{c}{[\AA]} & \multicolumn{1}{c}{[cts]} & \multicolumn{1}{c|}{[$10^{-5}\,\frac{\rm ph}{\rm s\,cm^2}$]} & \multicolumn{1}{c}{[\AA]} & \multicolumn{1}{c}{[\AA]} & \multicolumn{1}{c}{[cts]} & \multicolumn{1}{c|}{[$10^{-5}\,\frac{\rm ph}{\rm s\,cm^2}$]} & Ion & Trans. & $\lambda$ \\ \hline
\multicolumn{4}{c|}{\bf LETGS +1 order} \myrule & \multicolumn{4}{c|}{\bf LETGS -1 order} & \multicolumn{4}{c|}{\bf RGS\,1 + 2} & \\ \hline
$^*6.665$ & $0.035$ & $ 36.1 \pm ~8.8$ & $ 3.5 \pm ~0.8$ & $^*6.622$ & $0.027$ & $ 54.0 \pm ~8.8$ & $ 4.8 \pm ~0.8$ & $6.654$   & $0.039$ & $ 45.6 \pm 12.4$ & $ 1.8 \pm 0.5$ & Si\,XIII & r          & 6.650 \\  
$^*{\prime\prime}$ & ${\prime\prime}$ &  ${\prime\prime}$ & ${\prime\prime}$ & $^*{\prime\prime}$ & ${\prime\prime}$ & ${\prime\prime}$ & ${\prime\prime}$ & $6.729$ & $0.035$ & $ 41.1 \pm 11.7$ & $1.6 \pm 0.5$ & Si\,XIII & i          & 6.690 \\
$^*6.740$ & $0.039$ & $ 18.5 \pm ~8.0$ & $ 1.8 \pm ~0.8$ & $^*6.741$ & $0.033$ & $ 33.0 \pm ~7.8$ & $ 2.9 \pm ~0.7$ & ${\prime\prime}$ & ${\prime\prime}$ & ${\prime\prime}$ & ${\prime\prime}$ &  Si\,XIII & f          & 6.740 \\
$$        & $$      & $$               & $$             & $-$        & $-$               & $-$              & $-$   & $14.050$  & $0.044$ & $198.4 \pm 26.0$ & $ 2.8 \pm 0.4$ & Ni\,XIX & Ne8A & 14.045 \\
$$        & $$      & $$               & $$              & $14.208$  & $0.036$ & $ 37.8 \pm ~9.6$ & $ 4.9 \pm ~1.2$ & $14.197$  & $0.062$ & $352.5 \pm 39.4$ & $ 4.9 \pm 0.5$ & Fe\,XVIII & F14 & 14.208 \\
$14.241$  & $0.022$ & $ 45.5 \pm ~9.0$ & $ 6.0 \pm ~1.2$ & $$        & $$      & $$               & $$              & $14.252$  & $0.056$ & $183.7 \pm 35.6$ & $ 2.5 \pm 0.5$ & Fe\,XVIII & F13 & 14.257 \\
$-$       & $-$     & $-$              & $-$            & $-$        & $-$     & $-$             & $-$              & $14.379$  & $0.061$ & $212.2 \pm 29.8$ & $ 3.1 \pm 0.4$ & Fe\,XVIII & F12 & 14.378 \\
$-$       & $-$     & $-$              & $-$            & $-$        & $-$     & $-$             & $-$              & $14.525$  & $0.062$ & $274.2 \pm 30.6$ & $ 3.8 \pm 0.4$ & Fe\,XVIII & F10 & 14.540 \\
$15.036$  & $0.022$ & $107.9 \pm 12.1$ & $13.8 \pm ~1.6$ & $15.013$  & $0.021$ & $118.9 \pm 12.6$ & $14.9 \pm ~1.6$ & $15.017$  & $0.048$ & $961.0 \pm 41.5$ & $13.3 \pm 0.6$ & Fe\,XVII & Ne3C & 15.014 \\
$^*15.183$& $0.023$ & $ 18.4 \pm ~7.4$ & $ 2.3 \pm ~0.9$ & $^*15.193$& $0.032$ & $ 40.4 \pm ~9.7$ & $ 5.0 \pm ~1.2$ & $^*15.183$& $0.049$ & $282.9 \pm 31.9$ & $ 4.3 \pm 0.5$ & O\,VIII  & Ly$\gamma$ & 15.175 \\
$^*15.262$& $0.025$ & $ 47.4 \pm ~9.5$ & $ 6.0 \pm ~1.2$ & $^*15.271$& $0.017$ & $ 33.2 \pm ~8.0$ & $ 4.1 \pm ~1.0$ & $^*15.259$& $0.043$ & $270.3 \pm 30.2$ & $ 4.2 \pm 0.5$ & Fe\,XVII & Ne3D & 15.265 \\
$15.636$  & $0.002$ & $ 14.0 \pm ~4.7$ & $ 1.8 \pm ~0.6$ & $-$       & $-$     & $-$              & $-$             & $15.632$  & $0.022$ & $ 39.8 \pm 14.6$ & $ 0.6 \pm 0.2$ & Fe\,XVIII & F7 & 15.628 \\
$-$       & $-$     & $-$              & $-$             & $-$       & $-$     & $-$              & $-$             & $15.861$  & $0.037$ & $ 52.4 \pm 19.7$ & $ 0.8 \pm 0.3$ & Fe\,XVIII & F6 & 15.831 \\
$-$       & $-$     & $-$              & $-$ & $-$              & $-$& $-$              & $-$            &    ${\prime\prime}$ & ${\prime\prime}$ & ${\prime\prime}$ & ${\prime\prime}$ & Fe\,XVIII & F5 & 15.874 \\
$^*16.034$& $0.024$ & $ 95.6 \pm 11.5$ & $12.3 \pm ~1.5$ & $^*16.004$& $0.025$ & $104.0 \pm 11.8$ & $13.2 \pm ~1.5$ & $^*16.013$& $0.063$ & $770.1 \pm 45.8$ & $10.8 \pm 0.6$ & Fe\,XVIII & F4        & 16.002 \\
$^*{\prime\prime}$ & ${\prime\prime}$  & ${\prime\prime}$ & ${\prime\prime}$ & $^*{\prime\prime}$   & ${\prime\prime}$  & ${\prime\prime}$ & ${\prime\prime}$ & ${\prime\prime}$  & ${\prime\prime}$ & ${\prime\prime}$ & ${\prime\prime}$ & O\,VIII   & Ly$\beta$ & 16.003 \\
$^*16.110$& $0.020$ & $ 36.4 \pm ~8.2$ & $ 4.6 \pm ~1.0$ & $^*16.076$& $0.018$ & $ 19.3 \pm ~6.8$ & $ 2.4 \pm ~0.9$ & $^*16.080$& $0.066$ & $200.0 \pm 37.6$ & $ 2.8 \pm 0.5$ & Fe\,XVIII & F3 & 16.078 \\
$16.794$  & $0.017$ & $ 52.5 \pm ~8.7$ & $ 7.3 \pm ~1.2$ & $16.757$  & $0.031$ & $ 61.0 \pm 10.0$ & $ 7.7 \pm ~1.3$ & $16.783$  & $0.049$ & $575.9 \pm 32.9$ & $ 8.6 \pm 0.5$ & Fe\,XVII & Ne3F & 16.780 \\
$-$       & $-$     & $-$              & $-$             & $-$       & $-$     & $-$              & $-$             & $17.612$  & $0.019$ & $39.5 \pm 12.6$ & $ 0.6 \pm 0.2$ & Fe\,XVIII & F1  & 17.626 \\ 
$18.651$  & $0.029$ & $ 26.4 \pm ~6.9$ & $ 3.8 \pm ~1.0$ & $-$       & $-$     & $-$              & $-$             & $18.635$  & $0.064$ & $243.2 \pm 25.9$ & $ 3.6 \pm 0.4$ &O\,VII & He3A & 18.627 \\
$18.977$  & $0.023$ & $490.7 \pm 22.8$ & $71.4 \pm ~3.3$ & $18.947$  & $0.030$ & $484.0 \pm 22.8$ & $67.0 \pm ~3.2$ & $18.972$  & $0.059$ & $4630.1 \pm 74.0$ & $70.8 \pm 1.1$ &O\,VIII & Ly$\alpha$ & 18.973 \\
$-$       & $-$     & $-$              & $-$             & $-$       & $-$     & $-$              & $-$             & $28.460$  & $0.048$ & $ 72.6 \pm 15.8$ & $ 1.6 \pm 0.3$ & C\,VI  & Ly$\beta$  & 28.446 \\
$-$       & $-$     & $-$              & $-$             & $-$       & $-$     & $-$              & $-$             & $28.776$  & $0.021$ & $35.8 \pm 10.7$ & $~0.8 \pm ~0.2$ & N\,VI  & r          & 28.787 \\
$33.739$  & $0.017$ & $ 55.0 \pm ~8.4$ & $17.5 \pm ~2.7$ & $33.745$  & $0.026$ & $ 83.4 \pm 10.3$ & $23.3 \pm ~2.9$ & $33.732$  & $0.070$ & $662.7 \pm 31.1$ & $21.9 \pm 1.0$ & C\,VI  & Ly$\alpha$ & 33.700 \\
\hline
\multicolumn{4}{c|}{\bf LETGS +1 order} \myrule & \multicolumn{4}{c|}{\bf LETGS -1 order} & \multicolumn{4}{c|}{\bf RGS\,1} & \\ \hline
$21.614$  & $0.023$ & $ 82.4 \pm 10.2$ & $19.1 \pm ~2.4$ & $21.613$  & $0.022$ & $102.8 \pm 11.2$ & $21.8 \pm ~2.4$ & $21.600$  & $0.047$ & $475.0 \pm 26.0$ & $14.2 \pm 0.8$ & O\,VII & r & 21.610 \\
$21.812$  & $0.022$ & $ 19.8 \pm ~6.1$ & $ 4.6 \pm ~1.4$ & $21.810$  & $0.018$ & $ 17.6 \pm ~5.6$ & $ 3.8 \pm ~1.2$ & $21.814$  & $0.045$ & $ 99.0 \pm 15.3$ & $~3.0 \pm 0.5$ & O\,VII & i & 21.800 \\
$22.115$  & $0.016$ & $ 53.7 \pm ~8.3$ & $12.5 \pm ~1.9$ & $22.108$  & $0.022$ & $ 52.7 \pm ~8.5$ & $11.5 \pm ~1.9$ & $22.095$  & $0.044$ & $332.7 \pm 22.3$ & $10.1 \pm 0.7$ & O\,VII & f & 22.100 \\ 
\hline
\multicolumn{4}{c|}{\bf LETGS +1 order} \myrule & \multicolumn{4}{c|}{\bf LETGS -1 order} & \multicolumn{4}{c|}{\bf RGS\,2} & \\ \hline
$12.150$  & $0.021$ & $231.7 \pm 16.5$ & $32.5 \pm ~2.3$ & $12.130$  & $0.019$ & $169.8 \pm 14.4$ & $23.1 \pm ~2.0$ & $12.141$  & $0.044$ & $736.9 \pm 34.0$ & $21.2 \pm 1.0$ & Ne\, X   & Ly$\alpha$ & 12.132 \\
${\prime\prime}$ & ${\prime\prime}$ & ${\prime\prime}$ & ${\prime\prime}$ & ${\prime\prime}$   & ${\prime\prime}$  & ${\prime\prime}$ & ${\prime\prime}$ & ${\prime\prime}$& ${\prime\prime}$ & ${\prime\prime}$ & ${\prime\prime}$ & Fe\,XVII & Ne4C       & 12.134 \\
$12.291$  & $0.019$ & $ 33.5 \pm ~8.1$ & $ 4.7 \pm ~1.1$ & $12.266$  & $0.028$ & $ 10.0 \pm ~7.0$ & $ 1.4 \pm ~1.0$ & $12.282$  & $0.050$ & $140.1 \pm 21.6$ & $~4.0 \pm 0.6$ & Fe\,XVII & Ne4D & 12.264 \\
$-$       & $-$     & $-$              & $-$             & $12.839$  & $0.027$ & $ 22.9 \pm ~8.1$ & $ 3.1 \pm ~1.1$ & $12.836$  & $0.037$ & $63.3 \pm 15.5$ & $~1.8 \pm 0.5$ & Fe\,XX & N16 & 12.834 \\
$^*13.467$& $0.021$ & $127.4 \pm 13.1$ & $16.7 \pm ~1.7$ & $^*13.465$& $0.025$ & $129.3 \pm 13.7$ & $17.1 \pm ~1.8$ & $^*13.451$& $0.061$ & $545.8 \pm 35.9$ & $15.8 \pm 1.0$ & Ne\,IX & r & 13.448 \\
$^*13.535$& $0.024$ & $ 40.5 \pm  9.4$ & $ 5.3 \pm ~1.2$ & $^*13.543$& $0.030$ & $ 59.4 \pm 11.1$ & $ 7.8 \pm ~1.5$ & $^*13.539$& $0.072$ & $248.5 \pm 32.8$ & $~7.0 \pm 0.9$ & Fe\,XIX & O15 & 13.524 \\ 
$^{\prime\prime}$ & $^{\prime\prime}$ & $^{\prime\prime}$ & $^{\prime\prime}$ & $^{\prime\prime}$ & $^{\prime\prime}$ & $^{\prime\prime}$ & $^{\prime\prime}$ & $^{\prime\prime}$& $^{\prime\prime}$ & $^{\prime\prime}$ & $^{\prime\prime}$ & Ne\,IX & i & 13.553 \\ 
$13.720$  & $0.024$ & $ 71.9 \pm 10.6$ & $ 9.4 \pm ~1.4$ & $13.703$  & $0.033$ & $ 79.1 \pm 11.6$ & $10.3 \pm ~1.5$ & $^*13.709$& $0.068$ & $410.1 \pm 31.8$ & $11.3 \pm 0.9$ & Ne\,IX & f & 13.700 \\ 
$-$       & $-$     & $-$              & $-$             & $-$       & $-$     & $-$              & $-$             & $^*13.831$& $0.057$ & $74.3 \pm 21.6$ & $~2.0 \pm 0.6$ & Fe\,XVII & Ne3A & 13.826 \\
$17.083$  & $0.033$ & $157.4 \pm 14.2$ & $23.4 \pm ~2.1$ & $17.051$  & $0.035$ & $152.0 \pm 14.1$ & $20.8 \pm ~1.9$ & $^*17.050$& $0.053$ & $419.9 \pm 37.5$ & $12.1 \pm 1.1$ & Fe\,XVII & Ne3G & 17.055 \\
${\prime\prime}$ & ${\prime\prime}$ & ${\prime\prime}$ & ${\prime\prime}$ & ${\prime\prime}$ & ${\prime\prime}$ & ${\prime\prime}$ & ${\prime\prime}$ & $^*17.095$ & $0.052$ & $393.9 \pm 36.8$ & $11.4 \pm 1.1$ & Fe\,XVII & NeM2 & 17.100 \\
$24.802$  & $0.020$ & $ 34.3 \pm ~7.2$ & $ 8.0 \pm ~1.7$ & $24.810$  & $0.027$ & $ 57.5 \pm ~9.0$ & $12.7 \pm ~2.0$ & $24.788$  & $0.079$ & $331.6 \pm 23.4$ & $12.0 \pm 0.8$ & N\,VII & Ly$\alpha$ & 24.780 \\
\hline
\multicolumn{15}{l}{$^a$ Fluxes are not corrected for interstellar absorption.}
\end{tabular}
\end{center}
\end{sidewaystable*}

%
%
\begin{table*}\scriptsize
\begin{center}
\caption{Emission lines observed in the time averaged positive 
and negative LETGS spectra outside the spectral range of or not 
detected by the RGS. 
All lines are represented by Gaussians. Line identifications
are from \protect\citey{Mewe85.1}, \protect\citey{Mewe95.1}, and \protect\citey{Phillips99.1}.}
\label{tab:letgs}
\begin{tabular}{rrrr|rrrr|llr} \hline
\multicolumn{4}{c|}{\bf LETGS positive order} & \multicolumn{4}{c|}{\bf LETGS negative order} & & & \\
\multicolumn{1}{c}{$\lambda$} \myrule & \multicolumn{1}{c}{$\sigma$} & \multicolumn{1}{c}{$I$} & \multicolumn{1}{c|}{Photon Flux} & \multicolumn{1}{c}{$\lambda$} & \multicolumn{1}{c}{$\sigma$} & \multicolumn{1}{c}{$I$} & \multicolumn{1}{c|}{Photon Flux} & \multicolumn{3}{c}{Identification} \\
\multicolumn{1}{c}{[\AA]} \myrule & \multicolumn{1}{c}{[\AA]} & \multicolumn{1}{c}{[ctgs]} & \multicolumn{1}{c|}{[$10^{-5}\,\frac{\rm ph}{\rm s\,cm^2}$]} & \multicolumn{1}{c}{[\AA]} & \multicolumn{1}{c}{[\AA]} & \multicolumn{1}{c}{[cts]} & \multicolumn{1}{c|}{[$10^{-5}\,\frac{\rm ph}{\rm s\,cm^2}$]} & Ion & Trans. & $\lambda$ \\ \hline
$  6.195$  & $0.027$ & $ 36.2 \pm  7.6$ & $ 3.5 \pm 0.7$ & $  6.178$ & $0.026$ & $ 24.7 \pm  6.8$ & $ 2.3 \pm 0.6$ & Si\,XIV  & Ly$\alpha$ & 6.180 \\
$  8.437$  & $0.019$ & $ 18.7 \pm  6.4$ & $ 2.0 \pm 0.7$ & $  8.402$ & $0.019$ & $ 20.4 \pm  6.6$ & $ 2.1 \pm 0.7$ & Mg\,XII & Ly$\alpha$ & 8.421 \\ 
\hline 
$  36.424$ & $0.044$ & $ 38.2 \pm  7.6$ & $13.2 \pm 2.6$ & $36.394$ & $0.028$ & $ 18.4 \pm  5.5$ & $ 5.6 \pm  1.7$ & Ne\,X   & Ly$\alpha$(3.) & 36.396 \\
$^{\prime\prime}$ & $^{\prime\prime}$ & $^{\prime\prime}$ & $^{\prime\prime}$ & $^{\prime\prime}$ & $^{\prime\prime}$ & $^{\prime\prime}$ & $^{\prime\prime}$ & S\,XII  & B6A        & 36.398 \\
$-$        & $-$     & $-$              & $-$            & $36.559$ & $0.016$ & $  9.8 \pm  4.0$ & $ 3.0 \pm  1.3$ & S\,XII  & B6C        & 36.563 \\
$  36.912$ & $0.023$ & $ 15.3 \pm  4.9$ & $ 5.5 \pm 1.8$ & $-$      & $-$     & $-$              & $-$             & S\,X    & N7         & 36.900 \\
$ 37.954$ & $0.030$ & $ 28.0 \pm  6.4$ & $11.2 \pm 2.6$ & $ 37.934$ & $0.057$ & $ 24.3 \pm  6.9$ & $ 8.6 \pm 2.5$ & O\,VIII  & Ly$\alpha$ (2.) & \\
$ 44.014$ & $0.020$ & $ 30.0 \pm  6.3$ & $ 4.8 \pm 1.0$ & $ 44.024$ & $0.032$ & $ 32.5 \pm  7.6$ & $ 4.4 \pm 1.1$ & Si\,XII  & Li6A       & 44.020 \\
$ 44.176$ & $0.025$ & $ 35.1 \pm  6.9$ & $ 5.6 \pm 1.1$ & $ 44.169$ & $0.017$ & $ 30.5 \pm  6.8$ & $ 4.1 \pm 0.9$ & Si\,XII  & Li6B       & 44.165 \\
$ 56.920$ & $0.031$ & $ 41.4 \pm  7.4$ & $ 8.5 \pm 1.5$ & $-$       & $-$       & $-$              & $-$            & O\,VIII  & Ly$\alpha$ (3.) &  \\
$ 94.023$ & $0.059$ & $ 70.6 \pm  9.9$ & $28.8 \pm 4.0$ & $ 93.956$ & $0.052$ & $ 53.9 \pm  8.7$ & $23.3 \pm 3.8$ & Fe\,XVIII& F4A        & 93.930 \\
$108.450$ & $0.058$ & $ 42.1 \pm  8.2$ & $22.2 \pm 4.3$ & $108.418$ & $0.057$ & $ 34.0 \pm  7.6$ & $18.6 \pm 4.2$ & Fe\,XIX  & O6a        & 108.370 \\
$117.185$ & $0.058$ & $ 46.8 \pm  8.6$ & $26.0 \pm 4.8$ & $117.184$ & $0.085$ & $ 56.6 \pm  9.8$ & $31.9 \pm 5.5$ & Fe\,XXII & B11        & 117.170 \\
$128.818$ & $0.049$ & $ 33.4 \pm  7.3$ & $33.6 \pm 7.4$ & $128.738$ & $0.055$ & $ 27.7 \pm  7.2$ & $24.8 \pm  6.5$ & Fe\,XXI  & C6A        & 128.730 \\
$132.978$ & $0.068$ & $ 80.8 \pm 10.6$ & $81.7 \pm 10.7$ & $132.952$ & $0.078$ & $ 95.5 \pm 11.4$ & $83.6 \pm 10.0$ & Fe\,XX   & N6A        & 132.850 \\
\hline
\end{tabular}
\end{center}
\end{table*}

\subsection{He-like ions}\label{subsect:he_ions}

\begin{figure*}
\begin{center}
\parbox{18cm}{
\parbox{9cm}{\resizebox{9cm}{!}{\includegraphics{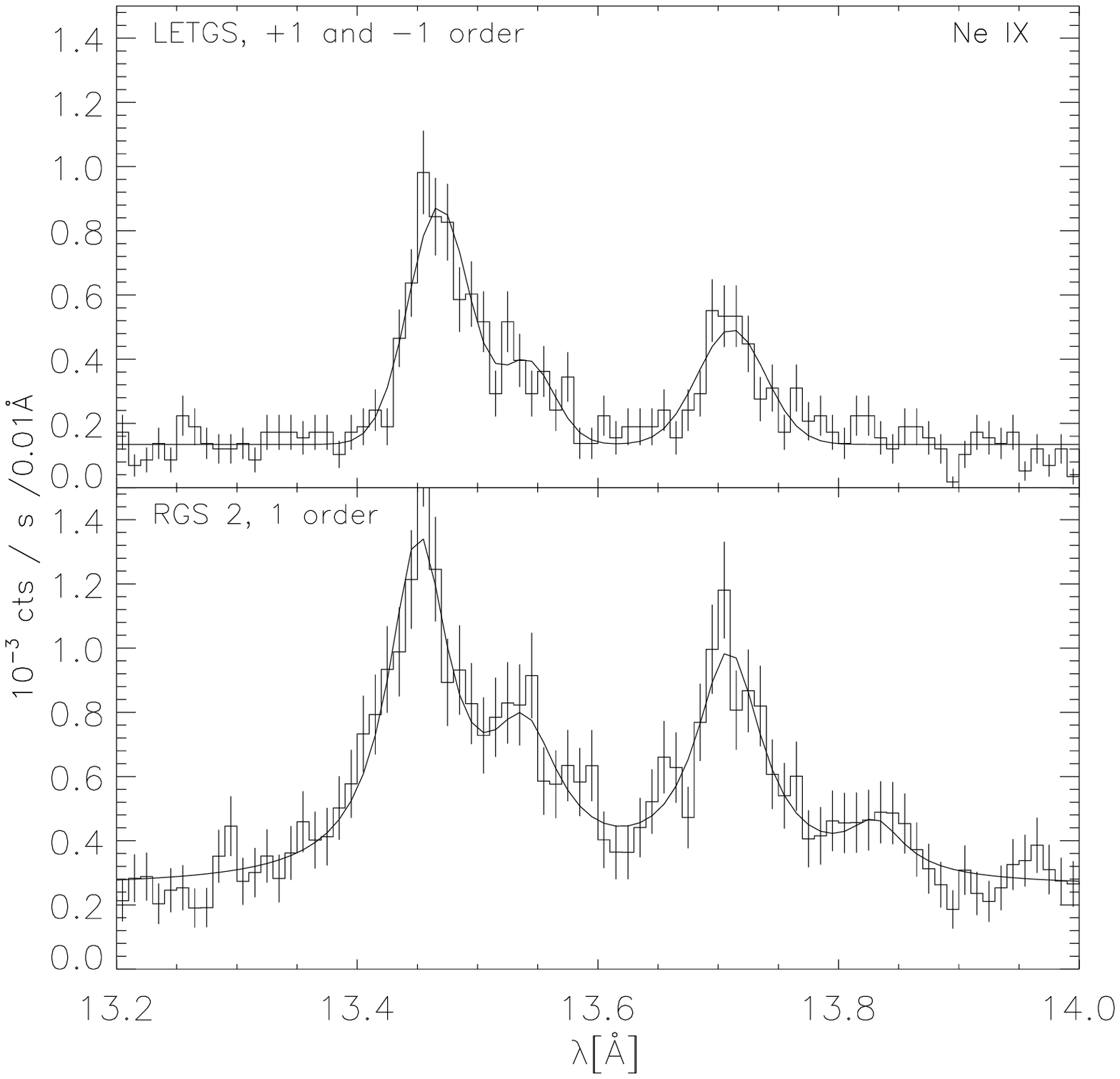}}}
\parbox{9cm}{\resizebox{9cm}{!}{\includegraphics{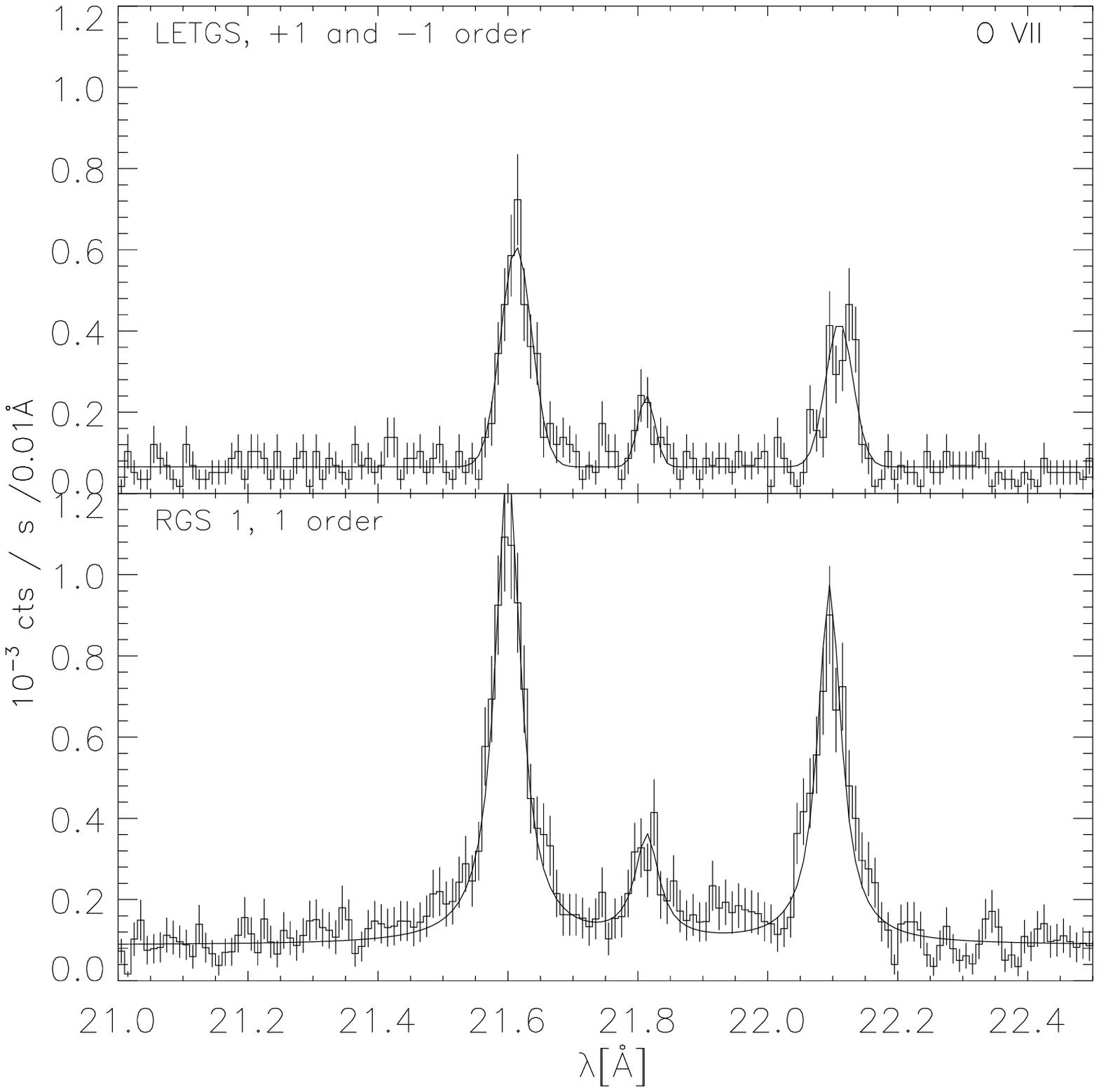}}}
}
\caption{He-like triplets of Ne\,IX and O\,VII 
measured with LETGS and RGS together with line fits. The continuum
close to each triplet is approximated by a straight horizontal line. 
The intensity enhancement to the right of the Ne\,IX triplet in the RGS 
spectrum is due to Fe\,XVII. The major contamination of the Ne\,IX triplet
comes from a Fe\,XIX line which is unresolved from the triplet 
intercombination line (see Table~\ref{tab:rgs_letgs}).}
\label{fig:triplets}
\end{center}
\end{figure*}

The spectral ranges of the LETGS and the RGS comprise  
the He-like triplet transitions of several elements. 
In the spectrum of \yygem the O\,VII triplet is by far the strongest
and best resolved triplet. 
The Si\,XIII triplet is not fully resolved, and the Ne\,IX triplet 
is blended with iron lines. All other
He-like ions do not produce sufficient signal.
The resonance line of the N\,VI triplet is weakly detected in the 
RGS spectrum, while the intercombination and the forbidden line of N\,VI 
are not identified. In Fig.~\ref{fig:triplets} we display 
the spectral region around the Ne\,IX and the O\,VII triplets. An iron line
on the long wavelength side of the Ne\,IX forbidden line 
(at $\lambda \sim 13.83$\,\AA) was included in the fit. 
The contamination of the Ne\,IX triplet stems mainly from Fe\,XIX
(at $13.53$\,\AA) which is unresolved from the triplet intercombination line.
Recall that for both the Ne\,IX and the O\,VII triplet only one RGS is available. 

%
%

The only element with resolved and strong enough lines for 
plasma diagnostics based on an 
analysis of line ratios is oxygen. 
Note that oxygen provides only information about 
comparatively low temperatures ($\sim 1-4$\,MK).
We determined the $R$ and $G$ ratios from the flux in the individual triplet 
lines. Furthermore, we computed the flux ratio between the H-like Ly$\alpha$ 
line and the He-like resonance line. This line ratio can be used as a 
temperature diagnostic, 
because the degree of ionization depends on the temperature. 
The result is listed in Table~\ref{tab:R_and_G}. 
%
%
%
%
%
\begin{table}
\begin{center}
\caption{Line ratios $R=f/i$, $G=(f+i)/r$, and $Ly\alpha/r$
for oxygen, 1\,$\sigma$ uncertainties.
}
\label{tab:R_and_G}
\begin{tabular}{lrrr}\hline
& \multicolumn{3}{c}{\bf LETGS} \\ 
& $+1$ & $-1$ & total \\ \hline
$R$                     &              $2.70 \pm 0.93$ & $3.01 \pm 1.08$ & $3.24 \pm 0.82$ \\
$G$                     &              $0.89 \pm 0.17$ & $0.70 \pm 0.13$ & $0.77 \pm 0.10$ \\
$Ly\alpha/r$            &              $3.73 \pm 0.49$ & $3.07 \pm 0.36$ & $3.35 \pm 0.29$ \\
\hline
& \multicolumn{3}{c}{\bf RGS} \\ \hline
$R$                     &             & & $3.38 \pm 0.57$ \\
$G$                     &             & & $0.92 \pm 0.08$ \\
$Ly\alpha/r$            &             & & $4.98 \pm 0.28$ \\
\hline
\end{tabular}
\end{center}
\end{table}
Fig.~\ref{fig:triplets} shows that  
the strength of the $r$ line is comparable to the sum of the intensities
in the $f$ and the $i$ line. This is typical for a collisional plasma.
It was shown by \citey{Blumenthal72.1} 
that the UV radiation field 
of the star can modify the $R$ ratio in the same way as collisional 
excitation, namely by depopulating the upper level of the $f$
transition to the upper levels of the $i$ transition.   
We made use of the calculations by \citey{Porquet01.1}, which
include the radiation field of the star, to estimate the 
density $n_{\rm e}$ and formation temperature $T$ of the
oxygen triplet. 
However, as \yygem is composed of two M-type stars it is
reasonable to neglect the radiation field and assume that the relative
intensities of $f$ and $i$ reflect the frequency of collisions and,
therefore, represent a direct measure of the density. 
The observed $G$ ratio corresponds to a formation temperature between 
$2$ and $3$\,MK.
The formation temperature defines the relation between the $R$ ratio and
the electron density (\cite{Porquet01.1}).
The models place our measurements of $R$ (see Table~\ref{tab:R_and_G}) in the
low-density limit, where $R$ is insensitive to density, and only an upper
limit of $n_{\rm e} \leq 2\,10^{10}\,{\rm cm^{-3}}$ can be provided.

The plasma emissivities from the MEKAL code 
(\cite{Mewe85.1} and \cite{Mewe95.1}) applied to the 
$Ly\alpha/r$ ratio result in a temperature of 
$3.4 \pm 0.1$\,MK in the LETGS and $4.0 \pm 0.1$\,MK in the RGS.
These somewhat higher temperatures than those 
derived from the $G$ ratio is naturally explained by the higher ionization
stage of the involved species.
The difference between the measurements in these two instruments 
probably arises from the difference in the O\,VII flux, which is 
underestimated in the RGS due to calibration problems as outlined above.
We caution that given these uncertainties the absolute values for the line
ratios must be treated with caution, but relative measurements comparing
the same instrument (e.g. as presented in Sect.~\ref{sect:timeres}) 
are still meaningful.

\subsection{Optical depth effects}\label{subsect:optdepth}

The general assumption of our plasma diagnostics is that the plasma
is optically thin, i.e., resonance photons are not scattered out of
the line of sight as described by
\citey{Schrijver95.1} and \citey{Mewe95.1}.
To check whether an optically thin model is justified 
we estimate the optical depth by comparing the observed flux ratio of the 
resonance line of Fe\,XVII at 15.014\,\AA\ and the adjacent line at 
15.265\,\AA\ of the same ion which has a much lower oscillator strength.  
The same procedure was carried out for other active stars, 
e.g. by \citey{Mewe01.1} and \citey{Ness02.1}, 
and they find $f_{\rm 15.014}/f_{\rm 15.265} = 2.85 \pm 0.14$
(Capella) and $f_{\rm 15.014}/f_{\rm 15.265} = 2.81 \pm 0.25$ (Algol).
For Proyon we derive a ratio of $2.9 \pm 1.7$ using the RGS measurement
given by \citey{Raassen02.1}. 
From comparison with laboratory measurements from EBIT 
($2.8-3.2$; \cite{Brown98.1}) \citey{Mewe01.1}
and \citey{Ness02.1} concluded that the coronal plasma of Procyon and Algol  
is optically thin. 
From our observation of YY\,Gem we calculate a ratio of 
$2.8 \pm 0.5$ for LETGS and $3.2 \pm 0.5$ for RGS, compatible with the
values cited above. This indicates that stellar coronae are indeed optically
thin, and as the results are similar for stars with very different activity
levels we conclude that optical depth effects generally are negligible in 
stellar coronae.

\section{Time-Resolved Spectroscopy \\ at High and Low Resolution}\label{sect:timeres}

In this section we complement the analysis of the emission line spectrum 
of \yygem with time-resolved modeling of the strongest lines seen with the 
RGS. The spectra discussed in the previous sections 
comprise the full {\em Chandra} and {\em XMM-Newton} 
exposure times and represent, therefore, 
averages over $\sim 75$\,\% of the orbital period of \yygem. 
The spectral shape is, however, expected to be variable (i) due to the
changing aspect 
of the binary in the course of its orbital motion, and (ii)
due to intrinsic variability related to activity on one or both of the
stars.

Next to a time-resolved analysis of individual lines, we study 
the temporal evolution of spectral parameters by a detailed 
inspection of the {\em XMM-Newton} EPIC spectrum.  
\citey{Guedel01.1} analysed the EPIC spectrum 
from the earlier {\em XMM-Newton} observation using MOS\,1 and MOS\,2 
individually for 5 different time intervals corresponding to different 
orbital phases of \yygem.
Here, we fit the pn and MOS\,1 + 2 spectra obtained in Sep 2000 
jointly. This allows to better constrain the spectral parameters
and perform a more detailed 
systematic analysis of time-resolved X-ray spectroscopy. 

To examine the evolution of the spectrum of \yygem we split the
total observing time in several time intervals.
In choosing the time bins we 
took into account the variability pattern in the lightcurve. 
We found that the LETGS data does not provide a strong enough signal
for time-resolved analysis, even if left and right side of the 
spectrum are added.
Therefore we make no further use of LETGS data for time-resolved  
spectroscopy.
The selection of the segments for EPIC and RGS 
can be read out of Fig.~\ref{fig:timeintervals}.

\subsection{The high-resolution RGS spectrum}\label{subsect:timeres_highres}

%
%
\begin{figure}
\begin{center}
\resizebox{9cm}{!}{\includegraphics{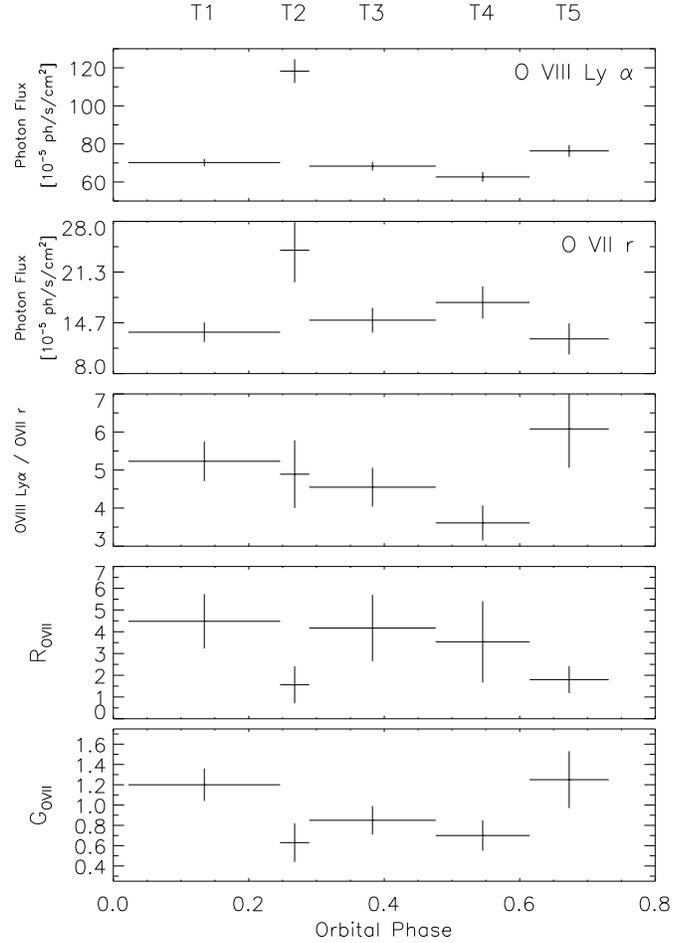}}
\caption{Time-evolution of the 
O\,VIII H-like Ly$\alpha$ line and the O\,VII He-like resonance
line, the ratio of these two lines, and the $R$ and $G$ ratio of the O\,VII triplet as observed by {\em XMM-Newton} RGS. Time intervals as defined in Fig.~\ref{fig:timeintervals}.}
\label{fig:rel_lineflux}
\end{center}
\end{figure}

We extracted RGS spectra in five time intervals 
(see Fig.~\ref{fig:timeintervals}).
Segments $2$ and $5$ are the phases with the highest level of activity.
The strongest lines in the RGS wavelength range are examined 
for all five selected data segments. 
Some of the weaker lines are not identified
at all phases. This may be due to variability of the source and/or 
the unequal S/N in the five intervals as a result of different exposure times.
The only element which displays a 
strong He-like triplet and H-like Ly$\alpha$ in all
time segments is oxygen. For this element we can refine the analysis of plasma
temperature and density deriving 
the $R$ and the $G$ ratio as a function of time.
It must be kept in mind, however, that oxygen probes only temperatures
between $1-4$\,MK, and therefore does not represent the major component
of the flare.

The spectra for the shorter time intervals are characterized by poor signal-to-noise. 
We ensured that the weak intercombination line is represented correctly
by holding the separation between the line centers within the triplet fixed
at the nominal value during the fitting process.
In Fig.~\ref{fig:rel_lineflux} we show the time evolution of
O\,VIII and O\,VII and their line ratios. 
All ratios are characterized by rather large error bars as a result of
short exposure times. 
A trend towards decreasing $R$ ratio during times of enhanced activity 
(intervals $2$ and $5$) is evident although only marginally significant.
To increase the significance of this result
we merged the three time intervals representing the quiescent state of
YY\,Gem, i.e., segments 
$1$, $3$, and $4$, and obtained an $R$ ratio of $3.91 \pm 0.79$.
The difference between this time-averaged quiescent $R$ value and
the value measured during segment $2$ is still only a 1.5\,$\sigma$ effect, 
but is suggestive for a density increase as a result either of 
chromospheric evaporation or compression of the coronal material.

\begin{figure*}
\begin{center}
\resizebox{18cm}{!}{\includegraphics{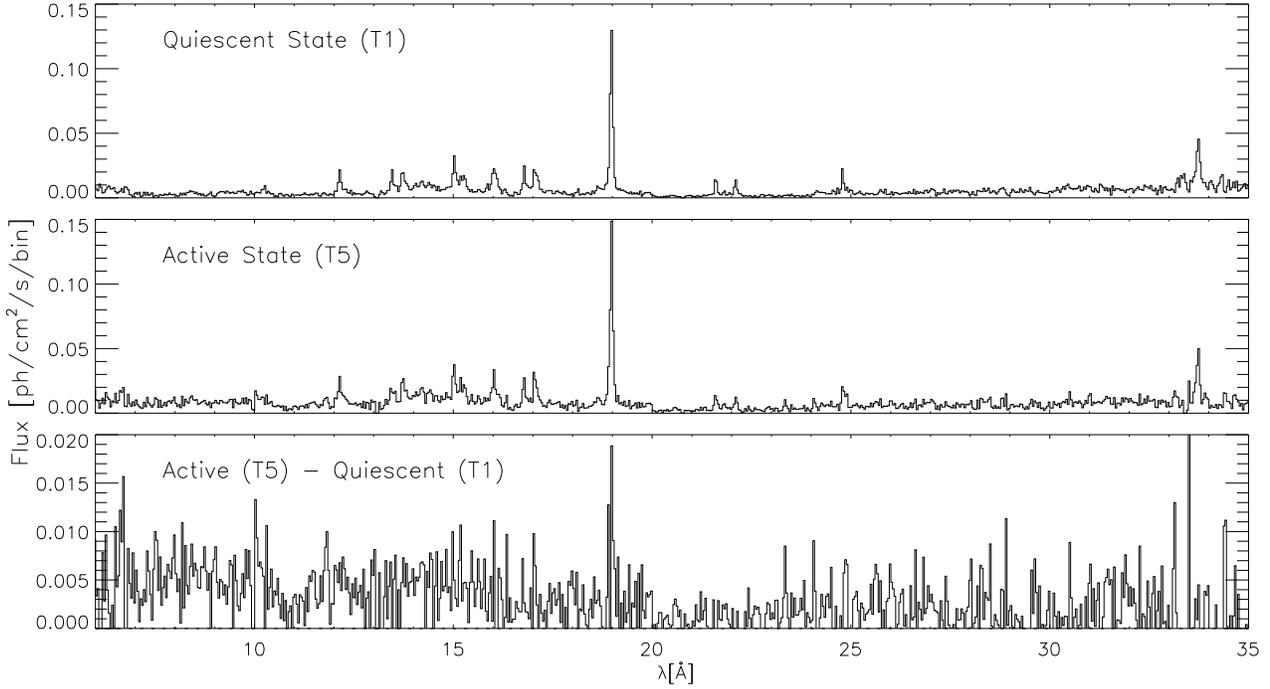}}
\caption{Comparison of the RGS flux spectrum 
during quiescence (interval $1$ from Fig.~\ref{fig:timeintervals}) and 
the most active phase (interval $5$ from Fig.~\ref{fig:timeintervals}). 
The panel on the bottom shows the difference spectrum with significant
excess emission below $\sim 16$\,\AA. Binsize is 0.04\,\AA.}
\label{fig:rgs_f_q}
\end{center}
\end{figure*}

In Fig.~\ref{fig:rgs_f_q} we compare 
the coadded RGS\,1 and RGS\,2 spectra during two different time intervals, 
segment $1$ from Fig.~\ref{fig:timeintervals} 
representing quiescence and 
segment $5$ from the same figure representing a phase of strong 
activity. The bottom panel of the figure shows that most of the flare 
emission is concentrated between $\sim 8-16$\,\AA. The low signal
does not permit to determine whether this emission is true continuum or 
whether it arises from unresolved iron lines which form at high temperatures
during the flare, e.g., Fe\,XXIV.
%
%

\subsection{{\em XMM-Newton} EPIC}\label{subsect:timeres_epic}

Due to its larger effective area the EPIC spectrum can be studied at
higher time-resolution. We split the EPIC pn and MOS 
lightcurve in a total of 15 time intervals 
(see Fig.~\ref{fig:timeintervals}). 

\subsubsection{Photon pile-up}\label{subsubsect:pile-up}

Photons arriving at the CCD detector can deposit energy in more
than one pixel. 
To improve the statistics in the high-energy portion 
of the spectrum we include double events, i.e., photons that produced
a signal in two adjacent pixels.

For bright sources (such as \yygem) 
more than one photon may arrive in one pixel 
before the detector is read out (photon pile-up), and in addition 
two photons arriving at the same time
in adjacent pixels are mistakenly identified as a single photon distributing
its energy on both pixels. 
Distortions of the spectral shape due to pile-up and mis-identification
of the pixel pattern distribution can safely be avoided by cutting out
the central portion of the point spread function, however, 
at the expense of count rate statistics. 
In order to compromise between correction for these 
effects and retaining enough events to constrain spectral parameters we
examined how model fits change if the photons from the innermost 
part of the source are excluded. 
For different inner radii of the source extraction area a 3-temperature
VMEKAL model (see Sect.~\ref{subsubsect:quies}) gives similar 
plasma temperature and abundances within the statistical uncertainties.
We conclude from this test 
that the spectral parameters are only marginally affected by pile-up
in this observation,
allowing us to use the full source region without having to 
remove the photons from the central area.

\subsubsection{The quiescent spectrum}\label{subsubsect:quies}

\citey{Guedel01.1} described the quiescent X-ray spectrum of \yygem
by a 3-temperature (3-T) model for thermal emission from an optically 
thin plasma. The existence of a multi-temperature corona on YY\,Gem
was verified by these authors also by deriving the emission measure
distribution, which showed the presence of a broad distribution of
temperatures ranging from $\sim 2-15$\,MK.
 
We used a combination of three VMEKAL-models 
as implemented in XSPEC 11.0.1 
to represent the spectrum in time interval $t_1$, when the source was
in quiescence.
The different instruments (pn, MOS\,1, and MOS\,2) were analysed jointly 
by multiplying the spectral model with a constant normalization 
factor to make up for uncertainties
in the absolute calibration of the individual detectors.

We used the EPIC spectrum of phase $t_1$ as baseline  
for all other time segments in which the signal is sometimes lower due to 
shorter exposure time. The model parameters for the quiescent spectrum
are provided in Table~\ref{tab:s2_v}. The EPIC spectrum shows that
besides the comparatively low temperatures probed by the strongest lines
in the high-resolution spectrum, plasma in excess of $10^7$\,K is present.
The observed spectrum can only be reproduced if the abundances of some
elements are free fit parameters. The best fit clearly shows subsolar
abundances, in particular for iron. Low abundances for elements with
low First Ionization Potential (FIP) were also reported from the earlier
{\em XMM-Newton} observation of YY\,Gem (see \cite{Guedel01.1}). 
\begin{table}
\begin{center}
\caption{Spectral parameters for the quiescent state of \yygem (derived
from time interval $t_1$). Normalization constants to cross-calibrate the
three instruments (pn, MOS\,1, MOS\,2) are: $N_{\rm pn} \equiv 1$ (fixed), 
$N_{\rm mos1}=1.01^{+0.03}_{-0.03}$, and $N_{\rm mos2}=1.03^{+0.03}_{-0.03}$.}
\label{tab:s2_v}
\begin{tabular}{rrrr}\hline\hline
\multicolumn{1}{c}{$kT_1$} & \multicolumn{1}{c}{$kT_2$} & \multicolumn{1}{c}{$kT_3$} & [${\rm keV}$] \\ \noalign{\smallskip} \hline \noalign{\smallskip}
$0.21^{+0.05}_{-0.07}$ & \myrule $0.64^{+0.01}_{-0.02}$ & $1.79^{+0.30}_{-0.24}$ & \\ \hline\hline
\multicolumn{1}{c}{$EM_1$} & \multicolumn{1}{c}{$EM_2$} & \multicolumn{1}{c}{$EM_3$} & [$10^{51}\,{\rm cm^{-3}}$] \\ \noalign{\smallskip} \hline \noalign{\smallskip} 
$2.24^{+1.87}_{-0.60}$ & \myrule $13.84^{+0.82}_{-2.88}$ & $2.88^{+1.05}_{-0.66}$ & \\ \noalign{\smallskip} \hline\hline \noalign{\smallskip} 
\multicolumn{1}{c}{O} & \multicolumn{1}{c}{Mg} & \multicolumn{1}{c}{Si} & \\ 
\noalign{\smallskip} \hline \noalign{\smallskip} 
$0.64^{+0.19}_{-0.14}$ & \myrule $0.27^{+0.12}_{-0.07}$ &$0.47^{+0.16}_{-0.07}$ & \\ \noalign{\smallskip} \hline\hline \noalign{\smallskip} 
\multicolumn{1}{c}{S} & \multicolumn{1}{c}{Fe} & \multicolumn{1}{c}{Ni} & $\chi^2_{\rm red}$ (dof) \\ \noalign{\smallskip} \hline \noalign{\smallskip} 
$0.50^{+0.27}_{-0.20}$ & $0.23^{+0.04}_{-0.03}$ & $0.00^{+0.28}_{-0.00}$ & 1.24 (658) \\ \noalign{\smallskip} \hline\hline \noalign{\smallskip} 
\end{tabular}
\end{center}
\end{table}

\subsubsection{Time-evolution of spectral parameters}\label{subsubsect:time-evolution}

At first we checked whether the spectrum of \yygem can 
be represented by a 3-T model at all times. We assumed that the 
integrated light from the quiescent corona remains unaltered during
the outburst. Therefore, we held all temperatures
and abundances fixed and allowed only the intensity to vary, corresponding
to changes in the emission measure as could be expected due to occultation
effects. This way we obtained acceptable fits
($\chi^2_{\rm red} \sim 1$ and statistically distributed residuals)
for all time intervals in which the star is `quiet', i.e. shows no strong
variability and exhibits intensity similar to the regular non-flaring state 
($t_2$, $t_7$, $t_8$, $t_9$, $t_{10}$, and $t_{12}$).
During the secondary eclipse (time intervals $t_8$ and $t_9$) 
the emission measures $EM_2$ and $EM_3$ drop because 
the visible fraction of the emitting corona has decreased. After the
eclipse a strong high-energy component (large $EM_3$)
is found. These time-intervals correspond to an emission feature in
the lightcurve (lasting from $t_{10}$ to $t_{12}$). For $t_{11}$,
representing the peak of this feature, we could not obtain an acceptable
3-T fit. The excess intensity in the hottest of the three components 
is seen already during the second half of the eclipse ($t_9$). 
This could be a hint that
the emission feature fills in part of the eclipse.

At all time intervals not discussed in the previous paragraph, 
after fitting the 3-T model
to the data a high-energy excess stands out in the residuals. This suggests
that higher temperature plasma is present in addition to the emission from
the quiescent corona. Adding a 4th VMEKAL component
does not lead to a significant improvement. Only a 5-T model represents
an adequate description of the data during the most 
active phases. 
A 5-T model was also introduced by \citey{Guedel01.1} to fit the
flare observed during the first {\em XMM-Newton} observation of \yygem.
For the modeling
we fixed spectral components \#\,$1-3$ at their quiescent values 
($t_1$; see Table~\ref{tab:s2_v}). 
Generally, the
statistics do not allow to constrain the abundances of the additional 
VMEKAL components. All abundances of components \#\,$4$ and \#\,$5$ 
were, therefore, held fixed at solar values. 
The `high state' ($t_{15}$) is an exception: Here 
the signal at the high-energy end of the EPIC spectrum is
better than for all other time-sliced spectra. 
Broad emission from the Fe K complex is clearly visible 
at $\sim$6.7\,keV, and an acceptable
solution is obtained only if the Fe abundance is a free parameter,
resulting in ${\rm Fe/Fe_\odot} = 0.47^{+0.10}_{-0.09}$. 

We point out that we performed an additional series of spectral fits 
where we set the abundances
of these components to the quiescent values, and obtain much larger 
uncertainties for the model parameters.
This hints at an abundance 
increase during active phases that seems to be not uncommon for 
stellar flares (\cite{Favata99.1}, \cite{Guedel99.1}). 

\begin{figure*}
\begin{center}
\parbox{\textwidth}{
\parbox{0.27\textwidth}{\resizebox{9cm}{!}{\includegraphics{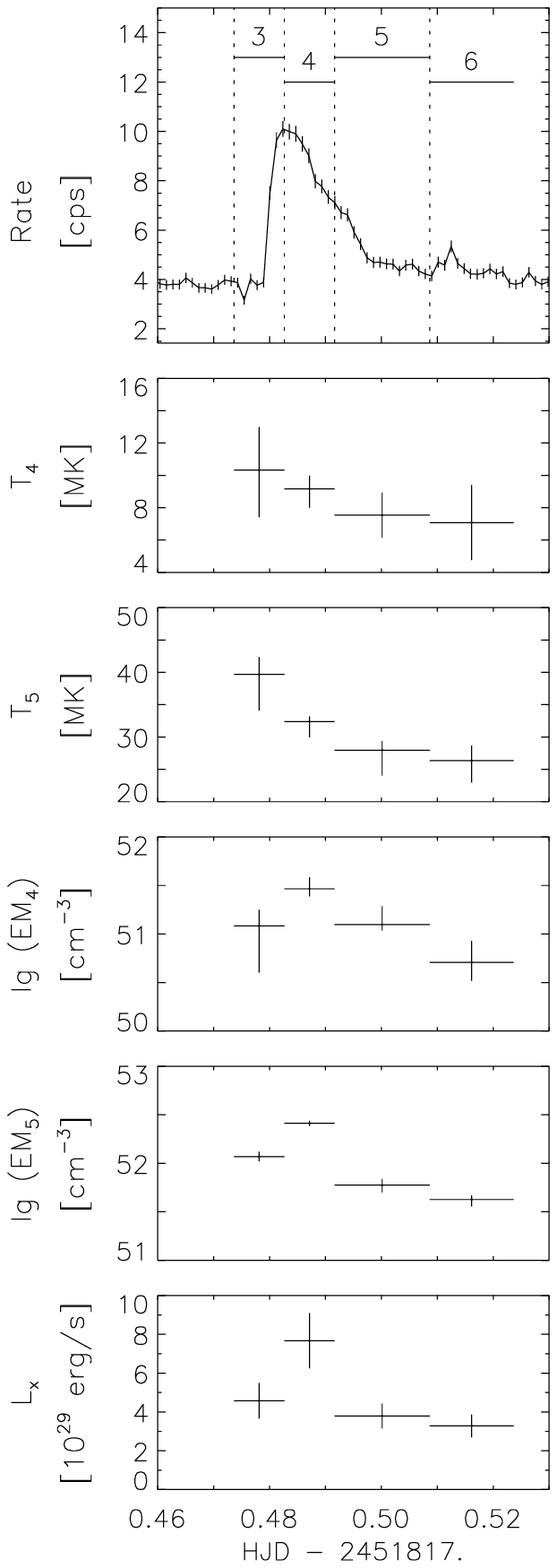}}}
\parbox{0.5\textwidth}{\resizebox{9cm}{!}{\includegraphics{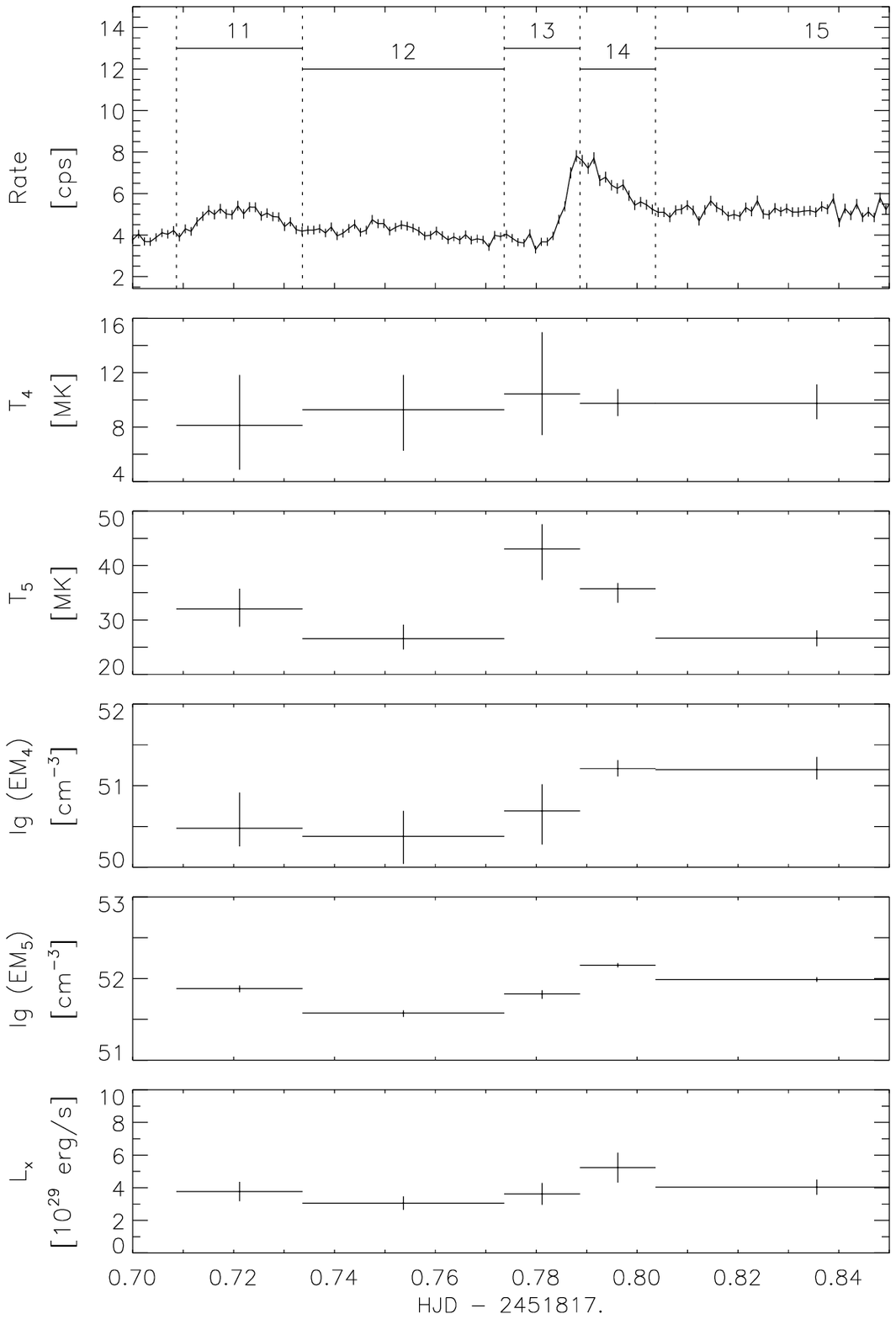}}}
}
\caption{Time evolution of spectral parameters during the two flares
on Sep 29/30, 2000. Shown are $T_4$, $T_5$, $EM_4$, and $EM_5$ from 5-T
model fits to the joint EPIC pn + MOS spectra of \yygem 
together with the pn lightcurve (top panel) and the
X-ray luminosity $L_{\rm x}$ (bottom panel). 
The remaining three spectral components were held fixed at 
their quiescent values. Error bars are 90\,\% confidence levels.}
\label{fig:flareparams}
\end{center}
\end{figure*}

Our modeling allows to investigate how the temperature and emission
measure of the flare components behave in different stages of flares. 
In Fig.~\ref{fig:flareparams} we
display both free temperatures and emission measures from the 5-T model 
as a function of time 
together with the EPIC pn lightcurve and the evolution of the luminosity
as derived from the spectral fits. Both spectral components show their highest
temperature during the rise phase of the lightcurve.
Subsequently, the temperatures decline
coherently with the intensity. The emission measure increases during the
rise phase and reaches its maximum with the beginning of the 
decay of the lightcurve. The observed luminosity $L_{\rm x}$ 
ranges from $(2 - 7.5)\,10^{29}\,{\rm erg/s}$ in the
EPIC band which is comparable to the values reported in the literature.
The spectral parameters during extended time intervals with 
enhanced count rate 
($t_{11}$, $t_{15}$) suggest that these are flare-like events
with a strong low-T component (VMEKAL \#\,4) and a weaker high-T component
(VMEKAL \#\,5).
But note that no typical flare outburst 
is seen in the lightcurve for the respective time intervals.

%

\section{Loop modeling}\label{sect:loops}

The size of emitting coronal structures can be estimated by help
of loop models. Most of these models are based on the 
assumption that the plasma is confined in a single loop. While this may
be adequate during simple flares, the situation in the quiescent corona
is likely to be much more complex. 

We used the temperature -- emission measure approach which allows
for heating during the flare decay, and 
determined the size of flaring loops from the spectral
parameters of the EPIC spectra. We restricted the analysis to the first
flare which is larger and longer than the second flare, and, therefore, 
provides better sensitivity for this kind of study. 

The evolution of temperature and emission measure puts important
constraints on the dynamics during flare decays. A one-dimensional 
hydrodynamic model describing the decay phase of solar flares was 
developed by \citey{Peres87.1} 
and discussed in a series of papers (see e.g., \cite{Serio91.1}, 
\cite{Jakimiec92.1}, \cite{Sylwester93.1}, \cite{Reale97.1}). 
The model was adapted to stellar flares by \citey{Reale97.1}. 
The duration of the heating (a free parameter in the model)
determines the slope $\zeta$ in the $\lg{T} - \lg{\sqrt{EM}} - $diagram. 
The limiting cases are (a) abrupt
switch off, and (b) a quasi steady-state evolution for slow decay of the
heating which leads to evolution along the path given by the RTV scaling
law (\cite{Rosner78.1}). 
The slope $\zeta$ and the decay time of the lightcurve $\tau_{\rm LC}$
are linked by the thermodynamic decay time scale, $\tau_{\rm td}$, i.e.
the decay time if heating were absent. 
This relation depends on instrumental properties. 
With the appropriate calibration $\zeta$ can be used to 
obtain an estimate for the half-length $L$ of the flare loop, since according
to Eq.~(2) in \citey{Reale97.1} $L$ is connected to the
maximum flare temperature, $T_{\rm max}$, 
and the thermodynamic decay time, $\tau_{\rm td}$.

\begin{figure}
\begin{center}
\resizebox{8.5cm}{!}{\includegraphics{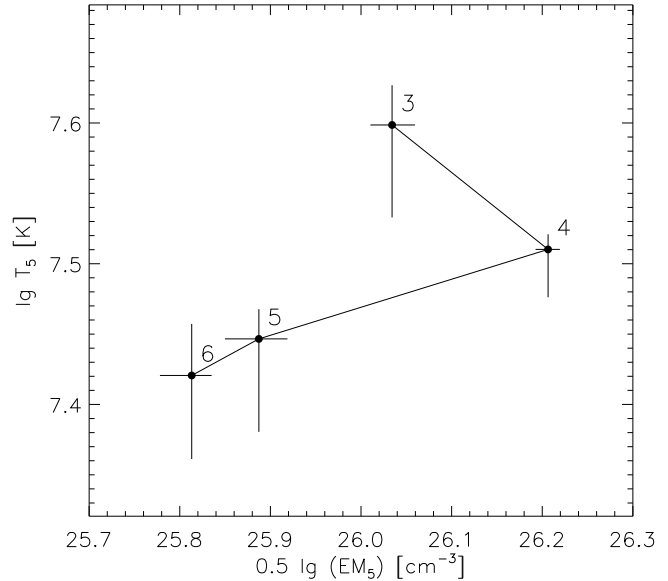}}
\caption{Temperature - emission measure - diagram for VMEKAL component \# 5 
for the larger of the two flares in the Sep 2000 {\em XMM-Newton} EPIC 
observation of \yygem. 
The numbers next to the data points indicate the respective time intervals.}
\label{fig:em_T}
\end{center}
\end{figure}

The $\lg{T} - \lg{\sqrt{EM}}$-diagram derived from our modeling of the 
{\em XMM-Newton} EPIC spectrum during the larger flare 
is shown in Fig.~\ref{fig:em_T}.
%
%
As the model was developed for a single-loop and one-temperature flare,
we ignored the first component (VMEKAL \# 4) of the flare spectrum.
This component has much smaller emission measure, i.e. the contribution 
to the flare emission is negligible compared to VMEKAL \# 5.
The curve in Fig.~\ref{fig:em_T} starts with the rise phase ($t_3$) 
of the flare. We measured the slope during the decay phase ($t_4$...$t_6$) 
and found: 
$\zeta=0.21$. 

We applied the model described above. 
Modeling the decaying part of the lightcurve by an exponential we
found $\tau_{\rm LC}=16 \pm 1$\,min.
The relation between $\zeta$ and $\tau_{\rm td}$ has been calibrated 
for EPIC-pn (Reale, priv. comm) and yields 
$L = 2.0\,10^9$\,cm for the length of the flare loop. 
As the low value we found for the slope $\zeta$ hints at continuous 
heating, the evolution of the plasma parameters can in approximation 
be thought of as quasi-static. Then, 
using the number for the loop length given above, the RTV scaling laws 
imply a pressure of 
$p \sim 6\,10^3\,{\rm dyne\,cm^{-2}}$ for a hydrostatic loop, 
and a heating rate of $E_{\rm H} \sim 45\,{\rm erg\,cm^{-3}\,s^{-1}}$.
This yields a density in the loop of $n \sim 6\,10^{11}\,{\rm cm^{-3}}$.
Assuming pressure equilibrium the density at temperatures typical for
the formation of O\,VII are estimated to be $\sim$\,10 times as high.
This contrasts with the low-density limit found from the analysis of the 
time-averaged O\,VII triplet, but is consistent with the high $R$ ratio found
during the flare. Therefore, the loop model underlines the indications for 
increased density during flares described in Sect.~\ref{sect:highres-spectra}.

\section{Summary and Conclusions}\label{sect:conclusions}

In September 2000 
\yygem was observed simultaneously by {\em XMM-Newton} and {\em Chandra}
providing a possibility to cross-check the performance of the 
instruments on both satellites. We found that the line centers and the
line fluxes measured by LETGS and RGS are in good agreement with only
few exceptions. One of these is the underestimation of the 
O\,VII resonance line flux due to an absorption edge in the RGS which was
not yet taken account of in the response matrix at the time of analysis  
(\cite{denHerder02.1}): For the resonance line of the O\,VII triplet
the RGS flux is found to be $\sim 70$\% of the LETGS flux,
but the measurements of the two instruments are 
compatible with each other at the 3\,$\sigma$ level. We note that 
similar differences
have been observed between LETGS and RGS for Procyon (see \cite{Raassen02.1}).
A deviation of the flux for the Ne\,X Ly$\alpha$ line between the two sides 
of the LETGS seems to point at a problem on the right side of this
detector at the respective wavelength (12.13\,\AA).

The high-resolution spectrum of \yygem is dominated by oxygen and neon
lines and lines from intermediate ionization stages of iron, generally 
Fe\,XVIII and below. 
Only few lines from highly ionized iron (Fe\,XIX and higher) 
are seen at the long wavelength end of the LETGS spectrum. 
The characteristic Fe\,XXI line at 128.73\,\AA~ is
detected but very weak, underlining that the corona of \yygem~ is not 
dominated by the high temperatures ($\geq 4\,$MK) typically observed on 
RS\,CVn binaries (see e.g., \cite{Mewe01.1}, \cite{Audard01.1}). 

We have probed the optical depth in the corona of YY\,Gem using
characteristic ratios of Fe\,XVII line fluxes. Comparing these ratios 
with measurements for other stars known to exhibit very different activity 
levels shows that optical depth effects do not
play a role in stellar coronae of all types. 
 
The correlation between X-ray luminosity and temperature 
in stellar coronae is well known (see e.g., \cite{Guedel97.1}).
Now, the dispersive instruments of {\em Chandra} and {\em XMM-Newton}
allow to confirm this relation using intensity ratios of individual
emission lines. 
\citey{Ness02.2} show in their comparison of {\em Chandra} observations of
ten stars that \yygem fits well into the
relation between $L_{\rm x}$ and temperature  
probed by the O\,VIII to O\,VII flux ratio  
representing a moderately active star.

The only element in the spectrum of YY\,Gem  
with a sufficiently strong and unblended helium-like triplet is oxygen.
The $R = f/i$ value derived from the observation presented here is compatible 
with an earlier measurement by \citey{Guedel01.1}. 
For the time-averaged spectrum the ratios of the oxygen triplet yield a
temperature of $T = 2-3$\,MK and an upper limit to the 
electron density of $n_{\rm e} \leq 2\,10^{10}\,{\rm cm^{-3}}$.
During the time of observation \yygem showed strong variability
including two flares. 
The time-resolved analysis of the high-resolution spectrum 
reveals variations of $R$ which are suggestive of an increase in density
during times of enhanced count rate. Our measurement is one of the first 
direct indications for the increase of the coronal density during 
flares on stars other than the Sun derived from high-resolution X-ray 
spectroscopy. A density measurement in the corona of AB\,Dor using an 
{\em XMM-Newton} measurement of oxygen triplet lines, e.g., 
resulted in comparable densities during flare and during quiescence 
(\cite{Guedel01.2}).
A detailed investigation of line ratios in the short time intervals 
representative for the different activity levels of \yygem is constricted 
by low S/N.
The time-resolved high-resolution spectroscopy, furthermore,
shows that during flares the absolute amount of (oxygen emission by) 
relatively low-temperature plasma increases. Lines from elements probing higher
temperatures more typical for flares were 
not strong enough for this kind of analysis. 

The presence of higher temperatures in the corona of \yygem is, however,
evident from our time-resolved spectroscopy with EPIC.
The temperatures for the global fit of the quiescent EPIC spectrum 
(2.4\,MK, 7.4\,MK, and 20.8\,MK) 
comprise the temperature range measured from the analysis
of oxygen lines, but go well beyond these values.
Two additional VMEKAL components are necessary to describe the 
EPIC spectrum during flares. 
The corresponding temperatures and emission measures 
are highest during the rise phase. 
An attempt to reproduce the EPIC spectrum with a number of 
Gaussian profiles using the information from the RGS 
spectrum was described by \citey{Haberl02.1}. But the low resolution of EPIC 
makes this instrument insensitive to changes 
in the relative strength of nearby emission lines during different phases of 
activity. 

With plasma parameters derived from spectral modeling 
the size of the (flare) emitting structures can be inferred. 
We used a hydrodynamic approach (see \cite{Reale97.1})  
involving heating during the decay phase  
to determine the length of flaring loops on YY\,Gem.
The overall size of quiescent stellar coronae can be estimated 
making use of plasma density and temperature derived from the analysis of 
high-resolution spectra (see e.g., \cite{Ness01.1}, 
\cite{Mewe01.1}, and \cite{Raassen02.1}). 
However, loop lengths derived from individual lines 
refer only to a very limited temperature range and the relevance of the
single loop model for the quiescent corona is questionable.
In our case the spectral parameters derived from individual lines are not
useful for this kind of analysis, since the upper limits on the density
translate to lower limits of the loop size according to 
the RTV-laws.
Instead, we derived loop sizes using results from global fitting 
with EPIC. 
We find that loops are small ($2\,10^9$\,cm), 
on the order of a few percent of the 
stellar radius ($R_* \sim 0.6\,R_\odot$) 
of each of the components in the YY\,Gem binary. 
This is consistent with the compactness of the corona established from 
eclipse mapping for the earlier {\em XMM-Newton} observation of YY\,Gem
(\cite{Guedel01.1}). Small loops point at
magnetic activity on the individual stars rather than large scale 
structures associated with overlapping magnetospheres.  
Furthermore, the RTV-laws point at high densities in the loop, 
confirming the evidences for low $R-$ratio during times of flares. 

\begin{acknowledgements}
BS wishes to thank D. Grupe and F. Haberl 
for helpful discussions on the RGS and EPIC data analysis.
We would like to thank the referee R. Pallavicini for helpful comments.
BS, VB, and RN are supported by the BMBF/DLR under grant numbers 
50~OR~0104, 50~OX~0001, and 50~OR~0003, respectively.
The PSI group acknowledges support from the Swiss National Science Foundation 
(grant 2100-049343).
JUN acknowledges financial support from DLR under 50~OR~98010, and 
NG acknowledges financial support from 
the European Union (HPMF-CT-1999-00228).
This work is based on observations obtained with {\em XMM-Newton}, 
an ESA science mission with instruments and contributions directly funded by 
ESA Member States and the USA (NASA).
\end{acknowledgements}

\end{document}